\documentclass[12pt, english]{uvamath}
\pdfoutput=1

\usepackage[english]{babel}
\usepackage{graphicx}
\usepackage[pdfborder={0 0 0}]{hyperref}
\usepackage{lipsum}
\usepackage{pgfplots}

\usepackage{amsmath}
\usepackage{amssymb}  
\usepackage{amsthm}
\usepackage{tikz}    
\usetikzlibrary{external, chains, shapes.geometric, decorations.pathmorphing}
\usepackage{enumitem}
\usepackage{subfigure}

\title{Satisfiability of Short Circuit Logic}
\author[sander.intveld@student.uva.nl]{Sander in 't Veld}

\what{Bachelor's Thesis in Mathematics and Computer Science (revised)}
\supervisors{dr.\ Inge Bethke, prof.\ dr.\ Jan van Eijck}
\coverimage{
\tikzsetnextfilename{voorpagina}
\begin{tikzpicture}[diamond, minimum size=30pt, scale=0.7]

\foreach \x in {-3,...,1} {
\foreach \y in {-3,...,0} {
\pgfmathtruncatemacro{\a}{80/(1+abs(\x))}
\node[fill=blue!\a] at (3+\x+\y, 3+\y-\x) {};
} }
\foreach \x in {-3,...,0} {
\foreach \y in {0,...,3} {
\pgfmathtruncatemacro{\a}{80/(1+abs(\y))}
\node[fill=purple!\a] at (3+\x+\y, 3+\y-\x) {};
} }
\foreach \x in {1,...,3} {
\foreach \y in {-3,...,1} {
\pgfmathtruncatemacro{\a}{160/(2+abs(\x)+abs(\y))}
\node[fill=purple!\a] at (3+\x+\y, 3+\y-\x) {};
} }
\foreach \x in {1,...,3} {
\foreach \y in {0,...,3} {
\pgfmathtruncatemacro{\a}{80/(1+abs(\y))}
\node[fill=red!\a] at (3+\x+\y, 3+\y-\x) {};
} }

\end{tikzpicture}
}

\institute{
Faculteit der Natuurwetenschappen, Wiskunde en Informatica\\\smallskip
Universiteit van Amsterdam\\\bigskip
\includegraphics[width=0.070\hsize]{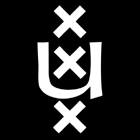}
}
\instituteaddress{Universiteit van Amsterdam\\
Science Park 904, 1098 XH Amsterdam\\
\url{http://www.science.uva.nl}{}
}

\theoremstyle{plain}                    
\newtheorem{stelling}{Theorem}[chapter]
\newtheorem{lemma}[stelling]{Lemma}  
\newtheorem*{lemma*}{Lemma}      
\newtheorem{propositie}[stelling]{Proposition} 
\newtheorem*{propositie*}{Proposition}            
\newtheorem{gevolg}[stelling]{Corollary}

\theoremstyle{definition}
\newtheorem{definitie}[stelling]{Definition}  
\newtheorem{voorbeeld}[stelling]{Example}

\theoremstyle{remark}
\newtheorem*{opmerking}{Remark}
\newtheorem*{bewijs}{Proof}

\setlist[enumerate]{label=\alph*., ref=(\alph*.)}

\newcommand{\field}[1]{\mathbb{#1}}
\newcommand{\R}{\field{R}}
\newcommand{\N}{\field{N}}

\newcommand{\A}{\mathcal{A}}
\newcommand{\true}{\mathrm{T}}
\newcommand{\false}{\mathrm{F}}
\newcommand{\B}{\{\true,\false\}}
\newcommand{\error}{\text{undefined}}

\newcommand{\bu}{\bullet}

\newcommand{\sand}{~
 \mathbin{\setlength{\unitlength}{1ex}
 \begin{picture}(1.4,1.8)(-.3,0)
 \put(-.6,0){$\wedge$}
 \put(-.45,-0.05){\textcolor{white}{\circle*{0.6}}}
 \put(-.45,-0.05){\circle{0.6}}
 \end{picture}
 }}

\newcommand{\sor}{~
 \mathbin{\setlength{\unitlength}{1ex}
 \begin{picture}(1.4,1.8)(-.3,0)
 \put(-.6,0){$\vee$}
 \put(-.45,1.35){\textcolor{white}{\circle*{0.6}}}
 \put(-.45,1.35){\circle{0.6}}
 \end{picture}
 }}

\newcommand{\sel}{\trianglelefteq}
\newcommand{\ser}{\trianglerighteq}
\newcommand{\seT}{\true}
\newcommand{\seF}{\false}
\newcommand{\depth}{\mathrm{depth}}
\newcommand{\comp}{\mathrm{cx}}

\newcommand{\SAT}{\mathbf{SAT}}
\newcommand{\FAL}{\mathbf{FAL}}
\newcommand{\fr}{\mathbf{fr}}
\newcommand{\rp}{\mathbf{rp}}
\newcommand{\cn}{\mathbf{cr}}
\newcommand{\mem}{\mathbf{mem}}
\newcommand{\st}{\mathbf{st}}
\newcommand{\SATfr}{\SAT_\fr}
\newcommand{\SATrp}{\SAT_\rp}
\newcommand{\SATcn}{\SAT_\cn}
\newcommand{\SATmem}{\SAT_\mem}
\newcommand{\SATst}{\SAT_\st}
\newcommand{\FALfr}{\FAL_\fr}

\newcommand{\Kfr}{\fr}
\newcommand{\Krp}{\rp}
\newcommand{\Kcn}{\cn}
\newcommand{\Kmem}{\mem}
\newcommand{\Kst}{\st}
\newcommand{\FSCL}{\mathrm{FSCL}}
\newcommand{\RPSCL}{\mathrm{RPSCL}}
\newcommand{\CSCL}{\mathrm{CSCL}}
\newcommand{\MSCL}{\mathrm{MSCL}}
\newcommand{\SSCL}{\mathrm{SSCL}}
\newcommand{\EqFSCL}{\mathrm{EqFSCL}}
\newcommand{\EqRPSCL}{\mathrm{EqRPSCL}}
\newcommand{\EqCSCL}{\mathrm{EqCSCL}}
\newcommand{\EqMSCL}{\mathrm{EqMSCL}}
\newcommand{\EqSSCL}{\mathrm{EqSSCL}}

\newcommand{\se}{\mathrm{se}}
\newcommand{\va}{\mathrm{va}}
\newcommand{\vacr}{\mathrm{cva}}
\newcommand{\last}{\mathrm{last}}

\newcommand{\vast}{\mathrm{sva}}
\newcommand{\snf}{f}
\newcommand{\constru}{\mathfrak{u}}
\newcommand{\compatname}{regular}

\newcommand{\tr}{:}
\newcommand{\concat}{\cdot}
\newcommand{\e}{\epsilon}
\newcommand{\PATHSATfr}{\mathbf{PathSat}_\mathit{fr}}
\newcommand{\PATHSATrp}{\mathbf{PathSat}_\mathit{rp}}
\newcommand{\PATHSATmem}{\mathbf{PathSat}_\mathit{mem}}
\newcommand{\PATHFALfr}{\mathbf{PathFal}_\mathit{fr}}
\newcommand{\PATHFALrp}{\mathbf{PathFal}_\mathit{rp}}
\newcommand{\PATHFALmem}{\mathbf{PathFal}_\mathit{mem}}
\newcommand{\print}{\diamond}
\newcommand{\cont}{\mathrm{cn}}

\newcommand{\PROPSAT}{\mathbf{PropSat}}
\newcommand{\bigoh}{\mathcal{O}}

\newcommand{\PT}{P^\true}
\newcommand{\PF}{P^\false}
\newcommand{\PS}{P^*}
\newcommand{\PL}{P^\ell}
\newcommand{\PC}{P^c}
\newcommand{\PD}{P^d}

\begin{document}
\maketitle

\begin{abstract}
The logical connectives typically found in programming languages are similar to their mathematical counterparts, yet different due to
their short-circuit behaviour -- when evaluating them, the second argument is only evaluated if the first argument is not sufficient to determine the result.
Combined with the possibility of side-effects, this creates a different type of logic called Short Circuit Logic.
A greater theoretical understanding of this logic can lead to more efficient programming and faster program execution.

In this thesis, formula satisfiability in the context of Short Circuit Logic is discussed.
A formal definition of evaluation based on valuation algebras is presented, alongside an alternative definition based on valuation paths.
The accompanying satisfiability and `path-satisfiability' are then proven to be equivalent, and an implementation of path-satisfiability is given.
Although five types of valuation algebras can be discerned, there are only three corresponding types of valuation paths.
From this, conclusions are drawn about satisfiability and side-effects; the manner in which side-effects alter truth values is relevant when analysing satisfiability, but the side-effects themselves are not.
\end{abstract}

\setcounter{tocdepth}{1}
\tableofcontents

\newpage
\thispagestyle{empty}
\mbox{}

\chapter{Introduction}
\label{ch_intro}

The field of logic deals with formulas and truths.
In propositional logic, a formula containing proposition letters $p$ and $q$ is said to be \emph{satisfiable} if each of the letters can be assigned a value, either true or false, such that the formula as a whole becomes true. For example, the formula $p \wedge \neg q$, which is read as ``$p$ and not $q$'', is satisfiable by taking $p$ to be true and $q$ to be false. On the other hand, the formula $p \wedge \neg p$ is not satisfiable in propositional logic, as $p$ cannot be simultaneously true and false.

Consider the following code fragment, written in C-like pseudocode.
\begin{verbatim}
integer n = 0
boolean a() { ... }
boolean b() { ... }

if ( a() && b() && !a() ) 
{
  print("Hello")
}
\end{verbatim}
We have one integer variable \texttt{n} and two functions \texttt{a()} and \texttt{b()} that take no arguments and return booleans.
Whether or not `Hello' is printed only depends on the value of \texttt{n}. However, it is possible that nothing will ever be printed, no matter what value we choose.
For instance, if \texttt{a()} simply always returns true, then \texttt{!a()} will always be false, and the line \texttt{print("Hello")} will never be reached.
In this case, \texttt{print("Hello")} is a piece of ``dead code''. Being able to detect dead code is of great interest to compilers and optimisers,
as the dead code is often the result of an error by the programmer, 
and since removing it reduces memory and cpu usage.
If we translate the if-clause \texttt{a() \&\& b() \&\& !a()} to the logical formula $a \wedge b \wedge \neg a$, then detecting dead code is similar to answering the question ``Is this formula satisfiable?''.
This is one of the many reasons why logicians and computer scientists seek a greater understanding of satisfiability.
\vspace{5mm}

When evaluating the formula $x \wedge y$, we usually first evaluate $x$ and $y$ separately. Then $x \wedge y$ is true if both $x$ and $y$ are true, and it is false if at least one of $x$ and $y$ is false. However, if $x$ is false then knowing this is enough to determine that $x \wedge y$ must also be false; the value of $y$ no longer needs to be considered. Computer programs can make use of this fact in what is called \emph{short-circuit evaluation}. 

Common programming languages such as C, Java and Haskell feature short-circuit evaluation in the form of the logical connectives \texttt{\&\&} and \texttt{||}.
A typical example of an expression using such a connective is
\[
\texttt{(n != 0) \&\& (x/n < 1)}
\]
where \texttt{n} and \texttt{x} are integer variables. Here the right-hand side of the expression, which features a relatively expensive division operation, will only be evaluated if the left-hand side evaluates to true. Besides being expensive, the division operation comes with the danger of `division by zero', which will result in a program crash on most platforms. Short-circuit evaluation in this case ensures that the expression will always return a value, as expected.

Also of relevance to logic in computer programs are \emph{side-effects}; the evaluation of a formula might change the state of the context in which it is evaluated. `Division by zero' could be considered an example of this, but its effect is so drastic that we will not further discuss it here. Instead, the assignment operator \texttt{=} as found in the C language provides a better example. The expression \texttt{(n = 55)} will assign the value $55$ to \texttt{n} and return true. Clearly, the evaluation of such an expression will affect the evaluation of later expressions containing \texttt{x}.

Detecting dead code is \emph{similar} to solving propositional satisfiability, but not the same.
In propositional logic, the formula $p \wedge q \wedge \neg p$ is unsatisfiable, but if we fill in the functions \texttt{a()} and \texttt{b()} from our code fragment as
\begin{verbatim}
boolean a() { return (n == 0) }
boolean b() { return (n = 55) }
\end{verbatim}
then the program would print `Hello'.
Thus, short-circuit evaluation and side-effects appear to be part of a different kind of logic.

\vspace{5mm}
In \emph{Short Circuit Logic} \cite{scl1}, the semantics of short-circuit evaluation and side-effects are described in more detail. A new type of logic, the short-circuit logic, is introduced, and the logics FSCL, RPSCL, CSCL, MSCL and SSCL are defined and axiomatised.

This thesis attempts to formally define what evaluation and satisfiability mean in the context of Short Circuit Logic, and suggests and implements a few methods to test the satisfiability of a formula with regards to these five logics. Relevant questions are:
\begin{itemize}
\item \emph{How does satisfiability for Short Circuit Logic differ from traditional satisfiability?}
\item \emph{How do different types of side-effect change satisfiability?}
\item \emph{Can short-circuit evaluation be utilised while testing satisfiability?}
\end{itemize}
The next chapter will be spent laying the groundwork, as well as summarizing a few results from \cite{scl1}. In Chapter~\ref{ch_evalsat}, we will formally define evaluation and satisfiability for Short Circuit Logic. An implementation for testing satisfiability will be discussed in Chapter~\ref{ch_implem}, but it will at first seem incompatible with the definitions from Chapter~\ref{ch_evalsat}. The gap between theory and implementation will be bridged in Chapter~\ref{ch_pathsat}, where we will define an alternative definition of satisfiability.  Finally, Chapter~\ref{ch_conclu} will reconsider the questions asked above.

\chapter{Preliminaries}
\label{ch_prelim}

\section{Notation}
\label{sect_notation}

Throughout this thesis, we will consider the left-sequential short-circuit versions of the connectives $\wedge$ and $\vee$ used in traditional logic. Here `left-sequential' means that the left-hand side is evaluated before the right-hand side, and short-circuit means that the right-hand side is only evaluated if the left-hand side is not enough to determine the result. 
We will follow notation featured in \cite{fourvalues} and \cite{scl1} and use the symbols $\sand$ and $\sor$ for these connectives. 
Additionally, the symbols $\true$ and $\false$ will be used for the truth values `true' and `false' respectively, and the symbol $\neg$ for logical negation.
Furthermore, the symbols $\sel$ and $\ser$ will be used to describe certain binary trees.

In earlier work, the connectives $\sand$ and $\sor$ are defined based on Hoare's conditional, $\_ \triangleleft \_ \triangleright \_$. In this thesis, we are not specifically interested in this conditional, and directly use the results from these works.

\section{Formulas}
\label{sect_formulas}

Whereas propositional logic considers formulas over a certain set $\Phi$ of proposition letters, Short Circuit Logic considers formulas over a set $\A$ of \emph{atoms}. The intuitive difference between proposition letters and atoms is that atoms can have side-effects. Throughout this thesis we will assume we have fixed a set $\A$ of atoms. The formulas of Short Circuit Logic are given by a few basic rules. 
First, the constants $\true$ and $\false$ are formulas, and each atom $a \in \A$ is a formula as well. Furthermore, if $x$ and $y$ are formulas, then so are $\neg x$, $x \sand y$ and $x \sor y$. More formally:

\begin{definitie} 
The \emph{formulas} over $\A$ are defined by the following grammar:
\[
x ::= \true ~|~ a ~|~ \neg x ~|~ x \sand x
\]
where $a$ ranges over $\A$.
\end{definitie} 

The two symbols $\false$ and $\sor$ seem to be missing from the above definition. Although adding them is possible, it would make induction proofs slightly less practical. Therefore, as is not uncommon in other fields of logic, we define $\false$ and $\sor$ as abbreviations:
\begin{align*}
\false :&= \neg \true, & x \sor y :&= \neg (\neg x \sand \neg y).
\end{align*}
We need brackets to indicate precedence in more complicated formulas. As an example, $\neg (x \sand y)$ is the negation of $x \sand y$, whereas $\neg x \sand y$ is the conjunction of $\neg x$ and $y$.

Throughout this thesis, we will make repeated use of induction to the complexity of formulas and other objects. This can be formalised with an adequate definition of complexity, such as the following:

\begin{definitie} 
Let $x$ be a formula. The \emph{complexity} of $x$ is defined recursively: 
\begin{align*}
\comp(\true) &= 0, \\
\comp(a) &= 0, \text{~~~~~~for each~} a \in \A, \\
\comp(\neg x_1) &= 1 + \comp(x_1), \\
\comp(x_1 \sand x_2) &= 1 + \max\{\comp(x_1), \comp(x_2)\}.
\end{align*}
\end{definitie} 

However, we will usually just remark that a proof is by induction and omit any formal inductive structure, for the sake of brevity. Lastly, we define what it means for a formula to be `constant-free'.

\begin{definitie}
A formula is called \emph{constant-free} if it contains neither $\true$ nor $\false$, i.e. if it is defined by the following grammar:
\[
x ::= a ~|~ \neg x ~|~ x \sand x
\]
where $a$ ranges over $\A$.
\end{definitie}

\section{Short Circuit Logics}
\label{sect_scl}

Logics identify certain formulas.
That is, if $x$ and $y$ are formulas, then some logics might consider $x$ and $y$ to be `the same'; not in the sense of their structure or complexity, but in the way that they behave as formulas. For instance, the formulas $\true$ and $\neg \false$ are very different in appearance, but both have the same semantical interpretation: `true'. If $x$ and $y$ are identified formulas, then so are $\neg x$ and $\neg y$, as well as $x \sand z$ and $y \sand z$ for any $z$, etcetera.

In \cite{scl1} and \cite{propscl}, five short-circuit logics are introduced: $\FSCL$, $\RPSCL$, $\CSCL$, $\MSCL$ and $\SSCL$. The names are abbreviations of ``free --'', ``repetition-proof --'', ``contractive --'', ``memorizing --'' and ``static short-circuit logic'' respectively. We will not go in detail about their definitions, but instead briefly discuss the intuitive differences between the five logics.

The logic $\FSCL$ is the least identifying short-circuit logic. As such, this logic describes only the most fundamental properties of the symbols $\true$, $\neg$ and $\sand$. This logic allows all types of side-effects.

In $\RPSCL$, atoms must retain their value when evaluated multiple times in a row; that is, if $a$ is true, then $a \sand a$ must also be true. The logic $\CSCL$ takes this a bit further and demands that only the first evaluation of two identical atoms can have a side-effect. Thus, in $\CSCL$, if $a \sand b$ is true, then so is $a \sand a \sand b$, because the second occurence of $a$ cannot have a side-effect that makes $b$ false.
In $\MSCL$, the effects and values of atoms are `memorised' entirely. This means that once $a$ have been evaluated to true, any further evaluations of $a$ must also lead to true and can have no further side-effects.

The logic $\SSCL$ is the most identifying and restrictive short-circuit logic. In this logic, there are no side-effects; or rather, the side-effect of an atom cannot actually affect what values later atoms will take. As such, the logic $\SSCL$ is equivalent to propositional logic.
This means that if we take a formula in $\SSCL$ and replace every $\true$ by $\top$, every $\sand$ by $\wedge$, every atom $a \in \A$ by a corresponding proposition letter $p \in \Phi$, etcetera, then evaluating the formula in $\SSCL$ is the same as assigning either `true' or `false' to each of the proposition letters in the translation.

If $E$ is an axiom system, i.e. a collection of axioms, then we write $E \vdash x = y$ if the logical statement ``$x = y$'' can be proven by using axioms from $E$ and logical tautologies.
An axiom system is \emph{sound} for a logic if every two formulas that are proven equal by the axioms, are identified by the logic.
An axiom system is \emph{complete} for a logic if every two formulas that are identified by the logic, can be proven equal by the axioms.
If an axiom system is both sound and complete for a certain logic, then it axiomatises this logic.

The logic $\FSCL$ is axiomatised by the system $\EqFSCL$, while $\RPSCL$ is axiomatised by $\EqRPSCL$, etcetera. These axiom systems can be found in the appendix. The soundness and completeness of each of the respective axiom systems is discussed in \cite{scl1}.

\section{Evaluation Trees}
\label{sect_evaltrees}

Binary trees are one of the most simple ways to emulate choice: starting at the root of a tree, we can go down either the left or the right branch. Once we have gone down either, we may encounter another choice, and after that yet more choices, until we eventually arive at a `leaf', where the journey down the tree ends. In Short Circuit Logic, we are interested in a specific type of trees.

\begin{definitie} 
The \emph{trees} over $\A$ are defined by the following grammar:
\[
X ::= \seT ~|~ \seF ~|~ X \sel a \ser X
\]
where $a$ ranges over $\A$.
\end{definitie} 

Figure~\ref{treeanotb} depicts the tree $(\seF \sel b \ser \seT) \sel a \ser \seF$. In these trees, the `choices' are atoms from our set $\A$, and our leaves are truth values. The supposed meaning is this: starting at the root, we encounter the atom $a$. If $a$ is true, then we descend down the left branch and encounter another atom: $b$. However, if $a$ is false, we take the right branch and we immediately arrive at a leaf: $\seF$. This is reminiscent of the short-circuit behaviour we are looking for.

\begin{figure}
\centering
\tikzsetnextfilename{treeanotb}
\begin{tikzpicture}[->,shorten >=1pt,auto,node distance=1.5cm,
  thick,main node/.style={fill=green!20,circle, inner sep=0pt, minimum size=20pt,draw,font=\sffamily\bfseries},
  end node/.style={fill=blue!20,draw,font=\sffamily\bfseries}]

	\node[main node] (1) {$a$};
	\node[end node] (1r) [below right of=1] {$\false$}; \path[->] (1) edge (1r);
	\node[main node] (1e) [below left of=1] {$b$}; \path[->] (1) edge (1e);
	\node[end node] (1ee) [below left of=1e] {$\false$}; \path[->] (1e) edge (1ee);
	\node[end node] (1er) [below right of=1e] {$\true$}; \path[->] (1e) edge (1er);
\end{tikzpicture}
\caption{A graphical depiction of the tree $(\seF \sel b \ser \seT) \sel a \ser \seF$.}
\label{treeanotb}
\end{figure}

To allow us define trees recursively, we will use substitution. Suppose we have a tree $X$ and we want to somehow `extend' this tree, then we can do this by replacing each of its leaves by new trees. Formally, we define a \emph{substitution} as follows:

\begin{definitie} 
Let $X$, $Y$ and $Z$ be trees. We define $X[\seT \mapsto Y, \seF \mapsto Z]$ as:
\begin{align*}
\seT[\seT \mapsto Y, \seF \mapsto Z] &= Y \\
\seF[\seT \mapsto Y, \seF \mapsto Z] &= Z \\
(X_1 \sel a \ser X_2)[\seT \mapsto Y, \seF \mapsto Z] &= X_1[\seT \mapsto Y, \seF \mapsto Z] \sel a \ser X_2[\seT \mapsto Y, \seF \mapsto Z]
\end{align*}
\end{definitie} 

Thus, if $X$, $Y$ and $Z$ are trees, then $X[\seT \mapsto Y, \seF \mapsto Z]$ is the tree $X$ where each $\seT$ leaf is replaced by $Y$ and each $\seF$ leaf by $Z$. As an example, the tree in Figure~\ref{treeanotb} can also be written as $(\seT \sel a \ser \seF)[\true \mapsto (\seF \sel b \ser \seT), \seF \mapsto \seF]$. Additionally, note that the substitution $[\seT \mapsto \seT, \seF \mapsto \seF]$ does not alter trees, whereas $[\seT \mapsto \seF, \seF \mapsto \seT]$ simply swaps the $\seT$ and $\seF$ leaves.

The real significance of trees is given by the following definition:

\begin{definitie} 
The \emph{short-circuit evaluation tree} of a formula $x$, denoted $\se(x)$, is defined as follows:
\begin{align*}
\se(\true) &= \seT \\
\se(a) &= \seT \sel a \ser \seF \\
\se(\neg x) &= \se(x)[\seT \mapsto \seF, \seF \mapsto \seT] \\
\se(x \sand y) &= \se(x)[\seT \mapsto \se(y), \seF \mapsto \seF]
\end{align*}
\end{definitie} 

\begin{opmerking}
The following equalities can be derived:
\begin{align*}
\se(\false) &= \seF \\
\se(x \sor y) &= \se(x)[\seT \mapsto \seT, \seF \mapsto \se(y)]
\end{align*}
They are not part of the definition, as $\false$ and $\sor$ are abbreviations.
\end{opmerking}

The tree depicted in Figure~\ref{treeanotb} is in fact the $\se$-tree of $a \sand \neg b$.
Note that as the atom $a$ appears before atom $b$ in the formula $a \sand \neg b$, it also appears earlier (i.e. higher) in the tree. However, not all atoms from a formula necessarily appear in the tree, as is apparent from $\se(\true \sor a) = \seT$. Still, $\se$-trees exactly represent the `behaviour' of formulas. This fact is proven in \cite{scl1} by Theorems 2.1.7, 3.2.2 and 3.5.2, and summarised as the following theorem:

\begin{stelling} \label{eqiffseeq} 
If $x$ and $y$ are formulas, then $\EqFSCL \vdash x = y \Longleftrightarrow \se(x) = \se(y)$.
\end{stelling} 

We will need a few more definitions throughout the following chapters, along with a small proposition.

\begin{definitie} 
Let $X$ be a tree. The \emph{depth} of $X$ is defined recursively: 
\begin{align*}
\depth(\seT) =
\depth(\seF) &= 0 \\
\depth(X_1 \sel a \ser X_2) &= 1 + \max\{\depth(X_1), \depth(X_2)\}.
\end{align*}
\end{definitie} 

\begin{definitie} 
A tree is called \emph{closed} by $\seT$ or $\seF$ if all of its leaves are $\seT$ or $\seF$ respectively. A tree is called \emph{open} if it is not closed.
\end{definitie} 

\begin{propositie} \label{overopentrees} 
Let $X$, $Y$ and $Z$ be trees. If $X$ is open and at least one of $Y$ and $Z$ is open, then $X[\seT \mapsto Y, \seF \mapsto Z]$ is open.
\end{propositie} 

\begin{bewijs}
Let $X$, $Y$ and $Z$ be trees such that $X$ and at least one of $Y$ and $Z$ is open.
Suppose $Y$ is open. Because $X$ is open, it contains at least one $\seT$ leaf. In $X[\seT \mapsto Y, \seF \mapsto Z]$, this leaf is replaced by $Y$, and therefore this new tree is open because $Y$ is open. If $Y$ is not open, $Z$ must be open. Because $X$ is open, it also contains at least one $\seF$ leaf, which is replaced by $Z$ in the new tree, and now $X[\seT \mapsto Y, \seF \mapsto Z]$ is open because $Z$ is open. \qed
\end{bewijs}

\begin{gevolg} \label{constantfreeopen} 
If $x$ is a constant-free formula, then $\se(x)$ is open.
\end{gevolg} 

\begin{bewijs}
Of course $\true$ is not constant-free. Clearly $\se(a)$ is open for all $a \in \A$. Let $\neg x$ be constant-free, then so is $x$. By induction we may assume that this means $\se(x)$ is open, and therefore $\se(\neg x) = \se(x)[\seT \mapsto \seF, \seF \mapsto \seT]$ is also open.
Let $x \sand y$ be constant-free, then so are $x$ and $y$, thus $\se(x)$ and $\se(y)$ are open. Now Proposition~\ref{overopentrees} tells us that $\se(x \sand y)$ is also open.
By induction, every constant-free formula has an open $\se$-tree. \qed
\end{bewijs}

\section{Normal Form}
\label{sect_normalform}

One final preliminary is the \emph{normal form}. This type of formula bridges the gap between formulas and $\se$-trees. Its definition is slightly more complex and is justified in \cite{scl1}.

\begin{definitie} 
Consider the following grammar, where $a$ ranges over $\A$:
\begin{align*}
P ::=&~~ \PT ~~|~~ \PF ~~|~~ \PT \sand \PS \\
\PT ::=&~~ \true ~~|~~ (a \sand \PT) \sor \PT \\
\PF ::=&~~ \false ~~|~~ (a \sor \PF) \sand \PF \\
\PS ::=&~~ \PC ~~|~~ \PD \\
\PL ::=&~~ (a \sand \PT) \sor \PF ~~|~~ (\neg a \sand \PT) \sor \PF \\
\PC ::=&~~ \PL ~~|~~ \PS \sand \PD \\
\PD ::=&~~ \PL ~~|~~ \PS \sor \PC
\end{align*}
A formula is in \emph{normal form} if it is defined by $P$ in this grammar. The formulas defined by $\PT$ are known as $\true$-terms; $\PF$ defines $\false$-terms, $\PL$ defines $\ell$-terms and $\PS$ defines $*$-terms. The formulas of the form $\PT \sand \PS$ are known as $\true*$-terms.
\end{definitie} 

In \cite{scl1}, a function $\snf$ is defined that maps each formula to a formula that is in normal form, and the following theorem (Theorem 3.2.2 in \cite{scl1}) is proved.

\begin{stelling} 
If $x$ is a formula, then $\EqFSCL \vdash x = \snf(x)$.
\end{stelling} 

As an example, the $\snf$-image of $a \sand \neg b$ is $\true \sand (((a \sand \true) \sor \false) \sand ((\neg b \sand \true) \sor \false))$. Note that the segment corresponding to the atom $a$ is $((a \sand \true) \sor \false)$, which closely mimics its $\se$-tree, $\seT \sel a \ser \seF$. To further highlight the connection between normal forms and $\se$-trees, we will prove the following corollary:

\begin{gevolg} 
If $x$ and $y$ are formulas, then
\[
\se(x) = \se(y) ~~\Longleftrightarrow~~
\EqFSCL \vdash x = y ~~\Longleftrightarrow~~
\EqFSCL \vdash \snf(x) = \snf(y).
\]
\end{gevolg} 

\begin{bewijs}
The first bi-implication is given by Theorem~\ref{eqiffseeq}. The second follows from the fact that for any $E$: if $E \vdash x = y$ and $E \vdash y = z$, then $E \vdash x =z$. \qed
\end{bewijs}

Thus, this corollary implies that for every formula there is a normal form equivalent that behaves the same, and any other normal form that behaves the same is identified with it by $\FSCL$ and all higher logics.
Another fundamental property normal forms have is that the three types of normal form ($\true$-term, $\false$-term and $\true*$-term) correspond directly to the three types of trees (closed by $\seT$, closed by $\seF$, open) seen earlier. This is given by the following proposition:

\begin{propositie} \label{normalformfundamental} 
Let $x$ be a formula.
\begin{enumerate}
\item If $x$ is a $\true$-term, then $\se(x)$ is closed by $\seT$.
\item If $x$ is a $\false$-term, then $\se(x)$ is closed by $\seF$.
\item If $x$ is a $\ell$-term or a $\true*$-term, then $\se(x)$ is open.
\end{enumerate}
\end{propositie} 

\begin{bewijs}
For (\textit{a.}), notice that if $x$ and $y$ are formulas and $\se(y)$ is closed by $\seT$, then so is $\se(x \sor y)$, as all $\seF$'s in $\se(x)$ are replaced by $\se(y)$. Since $\se(\true) = \seT$ is closed by $\seT$, it follows by a simple inductive proof that $\se$-trees of all $\true$-terms are closed by $\seT$. 
Similarly, for (\textit{b.}), if $\se(y)$ is closed by $\seF$, then so is $\se(x \sand y)$; this shows that $\se$-trees of $\false$-terms are closed by $\seF$.
We are left to show (\textit{c.}). 

Suppose $x$ is a $\true$-term and $y$ is a $\false$-term. If we write out $\se((a \sand x) \sor y)$, we end up with $\se(x) \sel a \ser \se(y)$. Similarly $\se((\neg a \sand x) \sor y) = \se(y) \sel a \ser \se(y)$. Because $\se(x)$ is closed by $\seT$ and $\se(y)$ is closed by $\seF$, the $\se$-tree of a $\ell$-term contains both a $\seT$ leaf and a $\seF$ leaf. Thus every $\ell$-term has an open $\se$-tree.

By Proposition~\ref{overopentrees}, the conjunctions and disjunctions added in the $\PC$ and $\PD$ rules keep $*$-terms open. Finally, if $x$ is a $\true$-term and $y$ a $*$-term, then $\se(x)$ contains a $\seT$ leaf and $\se(y)$ is open, so $\se(x \sand y) = \se(x)[\seT \mapsto \se(y), \seF \mapsto \seF]$ is also open. This means all $\true*$-terms have open $\se$-trees. \qed
\end{bewijs}

\chapter{Evaluation and Satisfiability}
\label{ch_evalsat}

In propositional logic, the evaluation of a formula depends entirely on which proposition letters are true, and which are not. Once we have assigned a truth value, either true or false, to each proposition letter $p \in \Phi$, the entire formula becomes either true or false. In Short Circuit Logic, the possibility of side-effects somewhat complicates this. Not only can atoms be true or not, but the evaluation of an atom can affect the evaluation of the atoms that come after it. This means that the value assigned to an atom cannot be fixed, but rather depends on what atoms have been evaluated before it. A possible way of defining evaluation for short-circuit logics would be to somehow keep track of the atoms evaluated, and assign a value to an atom based on this `evaluation history'. However, as formulas are not bounded in size, such a history-based definition would perhaps be unwieldly.

Instead, we use `valuations' to assign a truth value to each atom. These valuations can be points in a grid, nodes in a graph, etcetera; what they are exactly does not matter, as long as they assign truth values. Side-effects now become transitions between valuations. By moving from one valuation to another, any further atoms are now evaluated in the new valuation, with possibly a different truth values. Thus, a formula can no longer be evaluated as is, but is instead evaluated \emph{at} a certain valuation.

The structures that collect these valuations and the transitions between them, are called `valuation algebras'. The definition is based on the definition of valuation algebras for propositional algebra in \cite{propalg} and the definition of Hoare-McCarthy algebras in \cite{hoarealg}.

\section{Valuation Algebras}
\label{sect_algebras}

\begin{definitie} 
A \emph{valuation algebra} is a non-empty set $V$, whose elements are called \emph{valuations}, combined with two functions: the \emph{evaluation} $/ : \A \times V \to \{\true, \false\}$ and the \emph{derivative} $\bu : \A \times V \to V$.
\end{definitie}  

So, a valuation algebra is a triple $(V, /, \bu)$. Instead of the valuations themselves assigning truth values to atoms, we abstract away from what valuations \emph{really} are, and let the function $/$ assign these values for each valuation. The function $\bu$ describes the transitions between the valuations. We use infix notation for both $/$ and $\bu$. Also, if $a$ is an atom, then we speak of `the evaluation of $a$' as being the function $a/ : V \to \{\true, \false\}$, and `the derivative of $a$' being $a \bu: V \to V$. The reason $V$ must be non-empty is simple: we want to evaluate formulas, and to do so we need at least one valuation.

We often write a valuation algebra simply as $V$, and use the symbols $/$ and $\bu$ to implicitly refer to the evaluation and derivative associated with $V$. We should be cautious about this, however. It is worth noting that valuations are just points or worlds or states, that any set of points can be part of a valuation algebras, and that two valuation algebras can have the same set of valuations. What really defines a valuation algebra is its evaluation and derivative. Therefore, if $\mathfrak{u} = (V, /, \bu)$ is a valuation algebra, we shall sometimes emphasise that $/$ and $\bu$ belong to $\mathfrak{u}$ by considering them ``in $\mathfrak{u}$''. We could use subscripts for this, but this would make reading the various equations a bit tiresome.

To be able to evaluate formulas, instead of just atoms, we expand the definition.

\begin{definitie} 
Let $(V, /, \bu)$ be a valuation algebra. For each formula $x$, we define functions $x / : V \to \{\true, \false\}$, the \emph{evaluation} of $x$, and $x \bu : V \to V$, the \emph{derivative} of $x$, by extending the evaluation $a/$ and derivative $a\bu$ for atoms $a \in \A$, as follows:
\begin{align*}
\true/H &= \true & \true \bu H &= H \\
(\neg x)/H &= \neg (x/H) & (\neg x) \bu H &= x \bu H \\
(x \sand y)/H &= 
	\left\{ \begin{array}{ll}
	y/(x \bu H) 		& \text{if~} x/H = \true \\
	\false 			& \text{otherwise}
	\end{array} \right. 
	& 
	(x \sand y) \bu H &=
	\left\{ \begin{array}{ll}
	y \bu (x \bu H) 	& \text{if~} x/H = \true \\
	x \bu H 			& \text{otherwise}
	\end{array} \right.
\end{align*}
where $x, y$ are formulas and $H \in V$.
\end{definitie} 

\begin{opmerking}
The following equalities can be derived by for $\false$ and $\sor$:
\begin{align*}
\false/H &= \false & \false \bu H &= H \\
(x \sor y)/H &= 
	\left\{ \begin{array}{ll}
	\true & \text{if~} x/H = \true \\
	y/(x \bu H) & \text{otherwise}
	\end{array} \right. 
	& 
	(x \sor y) \bullet H &=
	\left\{ \begin{array}{ll}
	x \bu H & \text{if~} x/H = \true \\
	y \bu (x \bu H) & \text{otherwise}
	\end{array} \right.
\end{align*}
where $x, y$ are formulas and $H \in V$.
\end{opmerking}

The definitions concerning $\true$ and $\neg$ speak for themselves. In the definition of $(x \sand y)/$ and $(x \sand y)\bu$ the short-circuit nature shows; if $x$ evaluates to false, then $x \sand y$ immediately evaluates to false as well. The second part, $y$, is not evaluated and is skipped entirely, thus does not cause any side-effects.

Given a formula $x$, a valuation algebra $V$ and a specific valuation $H \in V$, we can now evaluate the formula $x$ in $H$ by using these definitions to calculate $x/H$. The rest of this section will be spent discussing a few properties of valuation algebras.

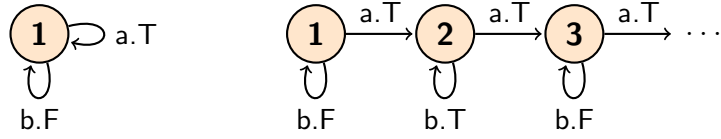
\begin{figure}
\centering
\subfigure{%
\tikzsetnextfilename{vavoorbeelda}
\begin{tikzpicture}[->,shorten >=1pt,auto,node distance=17mm, thick]
\tikzstyle{main node}=[circle,fill=orange!20,draw,font=\sffamily\bfseries]
\tikzstyle{end node}=[circle,fill=gray!10,draw,font=\sffamily\bfseries]

  \node[main node] (1) {1};

  \path[every node/.style={font=\sffamily\small}]
    (1) edge[loop right] node[right] {a.T} (1)
    (1) edge[loop below] node[below] {b.F} (1);
\end{tikzpicture}
}%
\hspace{10mm}
\subfigure{%
\tikzsetnextfilename{vavoorbeeldb}
\begin{tikzpicture}[->,shorten >=1pt,auto,node distance=17mm, thick]
\tikzstyle{main node}=[circle,fill=orange!20,draw,font=\sffamily\bfseries]
\tikzstyle{end node}=[circle,fill=gray!10,draw,font=\sffamily\bfseries]

  \node[main node] (1) {1};
  \node[main node] (2) [right of=1] {2};
  \node[main node] (3) [right of=2] {3};
  \node[draw=none] (5) [right of=3] {$\ldots$};

  \path[every node/.style={font=\sffamily\small}]
    (1) edge node[above] {a.T} (2)
    (2) edge node[above] {a.T} (3)
    (3) edge node[above] {a.T} (5)
    (1) edge[loop below] node[below] {b.F} (1)
    (2) edge[loop below] node[below] {b.T} (2)
    (3) edge[loop below] node[below] {b.F} (3);
\end{tikzpicture}
}%
\caption{Illustrations for two of the valuation algebras described in Example~\ref{algebraexamples}.}
\label{algebraexamplepics1}
\end{figure}

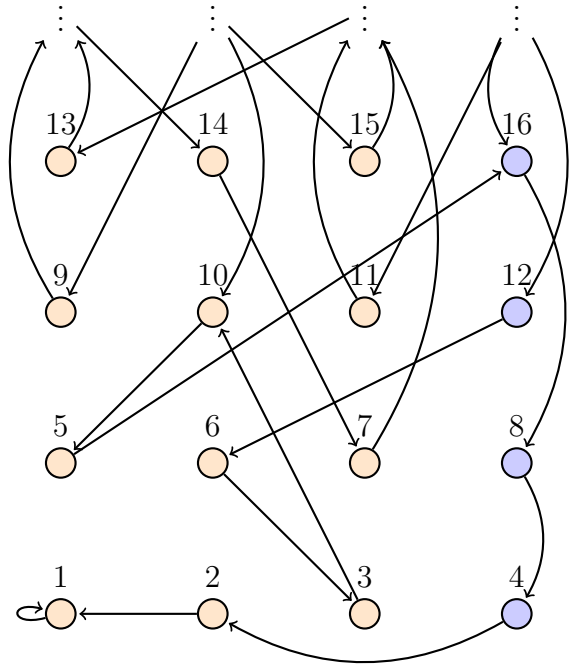
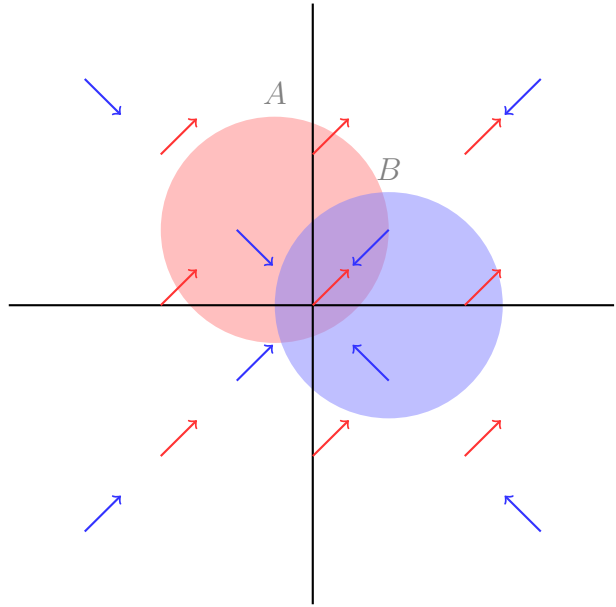
\begin{figure}
\centering
\subfigure[A segment of the Collatz tree. The arrows represent derivation by $\mathtt{a}$. The blue colour indicates where $\mathtt{b}$ is true. ]{%
\tikzsetnextfilename{vavoorbeeldc}
\begin{tikzpicture}[->,shorten >=1pt,auto, thick]
\tikzstyle{main node}=[circle, fill=orange!20, draw]
\tikzstyle{spec node}=[circle, fill=blue!20, draw]

\foreach \x in {1, ..., 3}
{
\foreach \y in {0, ..., 3}
{
	\pgfmathtruncatemacro{\label}{\x + 4 * \y};
	\node[main node, label=\label] (\label) at (2 * \x, 2 * \y) {};
}
\node[draw=none] (e\x) at (2 * \x, 8) {$\vdots$};
}
\foreach \y in {0, ..., 3}
{
	\pgfmathtruncatemacro{\label}{4 + 4 * \y};
	\node[spec node, label=\label] (\label) at (8, 2 * \y) {};
}
\node[draw=none] (e4) at (8, 8) {$\vdots$};

\path[loop left]	(1) edge (1);
\path 			(2) edge (1);
\path 			(3) edge (10);
\path[bend left] 	(4) edge (2);
\path		 	(5) edge (16);
\path 			(6) edge (3);
\path[bend right] 	(7) edge (e3);
\path[bend left] 	(8) edge (4);
\path[bend left] 	(9) edge (e1);
\path 			(10) edge (5);
\path[bend left] 	(11) edge (e3);
\path			(12) edge (6);
\path[bend right] 	(13) edge (e1);
\path			(14) edge (7);
\path[bend right] 	(15) edge (e3);
\path[bend left] 	(16) edge (8);
\path		 	(e2) edge (9);
\path[bend left] 	(e2) edge (10);
\path		 	(e4) edge (11);
\path[bend left] 	(e4) edge (12);
\path		 	(e3) edge (13);
\path		 	(e1) edge (14);
\path		 	(e2) edge (15);
\path[bend right] (e4) edge (16);
\end{tikzpicture}
}%
\hspace{10mm}
\subfigure[The sets $A$ and $B$ as a Venn diagram in $\R^2$. The arrows indicate the direction of derivation; red for $\mathtt{a}$, blue for $\mathtt{b}$.]{%
\tikzsetnextfilename{vavoorbeeldd}
\begin{tikzpicture}[shorten >=1pt,auto,thick, fill opacity=0.5]
\node[circle, minimum size =30mm, fill=red!50, label=$A$] (A) at (3.5, 5) {};
\node[circle, minimum size =30mm, fill=blue!50, label=$B$] (B) at (5, 4) {};

\path (4, 8) edge (4, 0);
\path (0, 4) edge (8, 4);

\path[->, draw=red!80] (2, 2) -> (2.5, 2.5);
\path[->, draw=red!80] (2, 4) -> (2.5, 4.5);
\path[->, draw=red!80] (2, 6) -> (2.5, 6.5);
\path[->, draw=red!80] (4, 2) -> (4.5, 2.5);
\path[->, draw=red!80] (4, 4) -> (4.5, 4.5);
\path[->, draw=red!80] (4, 6) -> (4.5, 6.5);
\path[->, draw=red!80] (6, 2) -> (6.5, 2.5);
\path[->, draw=red!80] (6, 4) -> (6.5, 4.5);
\path[->, draw=red!80] (6, 6) -> (6.5, 6.5);

\path[->, draw=blue!80] (3, 3) -> (3.5, 3.5);
\path[->, draw=blue!80] (3, 5) -> (3.5, 4.5);
\path[->, draw=blue!80] (5, 3) -> (4.5, 3.5);
\path[->, draw=blue!80] (5, 5) -> (4.5, 4.5);
\path[->, draw=blue!80] (1, 1) -> (1.5, 1.5);
\path[->, draw=blue!80] (1, 7) -> (1.5, 6.5);
\path[->, draw=blue!80] (7, 1) -> (6.5, 1.5);
\path[->, draw=blue!80] (7, 7) -> (6.5, 6.5);
\end{tikzpicture}
}%
\caption{Illustrations for two more valuation algebras described in Example~\ref{algebraexamples}.}
\label{algebraexamplepics2}
\end{figure}

In Example~\ref{algebraexamples}, a few valuation algebras are defined for two atoms, $\mathtt{a}$ and $\mathtt{b}$. It should be noted that a valuation algebra requires a properly defined evaluation and derivative function for \emph{all} atoms in $\A$. For practical reasons, we only show two. Figures~\ref{algebraexamplepics1}~and~\ref{algebraexamplepics2} depict the valuation algebras defined in the examples.

\begin{voorbeeld} \label{algebraexamples}
A few examples of valuation algebras.
\begin{enumerate}
\item 
The valuation algebra $(\{1\}, /, \bu)$ where $\mathtt{a}/1 = \true$, $\mathtt{b}/1 = \false$, $\mathtt{a} \bu 1 = 1$ and $\mathtt{b} \bu 1 = 1$.
\item
The valuation algebra $(\N, /, \bu)$ where $\mathtt{a}/n = \true$, $\mathtt{b}/n = \true$ if and only if $n$ is odd, $\mathtt{a} \bu n = n+1$ and $\mathtt{b} \bu n = n$ for $n \in \N$.
\item
The valuation algebra $(\N, /, \bu)$ where $\mathtt{a}/n = \true$ if and only if $n > 1$, $\mathtt{b}/n = \true$ if and only if $n$ is a multiple of 4, 
\[
\mathtt{a} \bu n =  \left \{ \begin{array}{ll}
	n/2 & \text{if $n$ is even} \\
	n & \text{if $n = 1$} \\
	3 \cdot n + 1 & \text{otherwise}
	\end{array} \right.
\]
and $\mathtt{b} \bu n = n$ for $n \in \N$.
\item
The valuation algebra $(\R^2, /, \bu)$ for some fixed sets $A \subseteq \R^2$ and $B \subseteq \R^2$, where $\mathtt{a}/(t_1, t_2)$ if and only if $(t_1, t_2) \in A$ and $\mathtt{b}/(t_1, t_2)$ if and only if $(t_1, t_2) \in B$, and where $\mathtt{a} \bu (t_1, t_2) = (t_1 + \frac{1}{3}, t_1 + \frac{1}{3})$ and $\mathtt{b} \bu (t_1, t_2) = (t_1 / 2, t_2 / 2)$.
\end{enumerate}
\end{voorbeeld}

The valuation algebra described in Example~\ref{algebraexamples}a only has one valuation, which means that there can be no side-effects. Therefore, evaluating a formula in this valuation algebra is similar to evaluating it in propositional logic, i.e. assigning either `true' or `false' to each atom in $\A$ and then resolving the formula. As such, these types of valuation algebras are not very interesting to us.

\begin{definitie} 
A valuation algebra that contains only one valuation is called \emph{trivial}.
\end{definitie} 

The valuation algebra from Example~\ref{algebraexamples}b is more interesting; it can be thought of as a program fragment based on a positive integer \texttt{n}, with two functions
\begin{verbatim}
boolean a()
{
  n = (n + 1)
  return true
}
boolean b()
{
  return (n % 2 == 0)
}
\end{verbatim}
where the C-like \texttt{n \% 2} returns 0 if \texttt{n} is even, and 1 if \texttt{n} is odd.

The third and fourth valuation algebras in Example~\ref{algebraexamples} are even more complex. In fact, it is not hard to imagine that there are practically no limits when it comes to `inventing' new valuation algebras, as long as the evaluation and derivative functions are properly defined. The range of valuation algebras is too wild and too expansive to accurately describe in four or five examples. Instead, we will characterise them by their properties.

\begin{definitie} 
Let $V$ be a valuation algebra, then $V$ is called
\begin{itemize}
\item \emph{repetition-proof} if $a/(a \bu H) = a/H$ for all $a \in \A$ and $H \in V$.
\item \emph{contractive} if $V$ is repetition-proof and $a \bu a \bu H = a \bu H$ for all $a \in \A$ and $H \in V$.
\item \emph{memorizing} if $V$ is contractive and 
\begin{align*}
a/(b \bu a \bu H) &= a/H, &
a \bu b \bu a \bu H &= b \bu a \bu H
\end{align*}
for all $a, b \in \A$ and $H \in V$.
\item \emph{static} if $V$ is memorizing and $a/(b \bu H) = a/H$ for all $a,b \in \A$ and $H \in V$.
\end{itemize}
\end{definitie} 

We will denote the collection of all valuation algebras by $\Kfr$, which stands for `free'. Moreover, we define $\Krp$, $\Kcn$, $\Kmem$ and $\Kst$ as the collections of repetition-proof, contractive, memorizing and static valuation algebras respectively. Note that $\Kst$ is a subcollection of $\Kmem$, which is a subcollection of $\Kcn$, etcetera.

From the names alone, one might suspect a link between the five collections of valuation algebras and the five short-circuit logics. The link is this: if two formulas are identified by, say, $\MSCL$, then they `behave' the same under all memorizing valuation algebras. To show this link, we will first need to properly define what it means to behave the same. We will define a relation called `valuation congruence' for each valuation algebra, and we will prove that this relation is in fact a congruence.

\begin{definitie} 
Let $V$ be a valuation algebra. Two formulas $x$ and $y$ are called \emph{valuation congruent} with respect to $V$ if $x/H = y/H$ and $x \bu H = y \bu H$ for all $H \in V$. We denote this by $x \equiv_V y$.
\end{definitie} 

\begin{propositie} \label{valconiscon} 
Let $V$ be a valuation algebra, then $\equiv_V$ is a congruence, i.e., for all formulas $x$, $y$, $x'$ and $y'$, if $x \equiv_V x'$ and $y \equiv_V y'$, then $\neg x \equiv_V \neg x'$ and $x \sand y \equiv_V x' \sand y'$.
\end{propositie} 

\begin{bewijs}
Let $V$ be a valuation algebra and let $x, y, x', y'$ be formulas such that $x \equiv_V x'$ and $y \equiv_V y'$. Using the definitions of evaluation and derivative, we find
\[
(\neg x)/H = \neg (x/H) = \neg (x'/H) = (\neg x')/H
\]
and
\[
(\neg x) \bu H = x \bu H = x' \bu H = (\neg x') \bu H
\]
for all $H \in V$. This means $\neg x \equiv_V \neg x'$. 

Because $x \bu H = x' \bu H$ for all $H$, we get $y/(x \bu H) = y/(x' \bu H)$, and since $y/G = y'/G$ for all $G$, including $G = x' \bu H$, we get $y/(x' \bu H) = y'/(x' \bu H)$. Also, because $x/H = \false \Leftrightarrow x'/H = \false$, we find
\begin{align*}
(x \sand y)/H
&= 	\left\{ \begin{array}{ll}
	y\phantom{'}/(x\phantom{'} \bu H) 	& \text{if~} x\phantom{'}/H = \true \\
	\false 								& \text{otherwise}
	\end{array} \right. \\
&= 	\left\{ \begin{array}{ll}
	y'/(x' \bu H) 		& \text{if~} x'/H = \true \\
	\false 			& \text{otherwise}
	\end{array} \right. \\
&= (x' \sand y')/H.
\end{align*}
Similarly, because $y \bu G = y' \bu G$ for all $G$, including $G = x' \bu H$, we get
\begin{align*}
(x \sand y) \bu H
&= 	\left\{ \begin{array}{ll}
	y\phantom{'} \bu (x\phantom{'} \bu H) 	& \text{if~} x\phantom{'}/H = \true \\
	x\phantom{'} \bu H 						& \text{otherwise}
	\end{array} \right. \\
&= 	\left\{ \begin{array}{ll}
	y' \bu (x' \bu H) 		& \text{if~} x'/H = \true \\
	x' \bu H 				& \text{otherwise}
	\end{array} \right. \\
&= (x' \sand y') \bu H
\end{align*}
and this proves $x \sand y \equiv_V x' \sand y'$. \qed
\end{bewijs}

The following theorem provides the desired connection between the five short-circuit logics and the five collections of valuation algebra. It is not proved in this thesis, but it is based on results proved in \cite{hoarealg} and to a lesser extent \cite{scl1}.

\begin{stelling}\label{equivissound} 
Let $x$ and $y$ be formulas.
\begin{enumerate}
\item $\EqFSCL \vdash x = y$ ~$\Longleftrightarrow$~ $x \equiv_V y$ for all $V$ in $\Kfr$.
\item $\EqRPSCL \vdash x = y$ ~$\Longleftrightarrow$~ $x \equiv_V y$ for all $V$ in $\Krp$.
\item $\EqCSCL \vdash x = y$ ~$\Longleftrightarrow$~ $x \equiv_V y$ for all $V$ in $\Kcn$.
\item $\EqMSCL \vdash x = y$ ~$\Longleftrightarrow$~ $x \equiv_V y$ for all $V$ in $\Kmem$.
\item $\EqSSCL \vdash x = y$ ~$\Longleftrightarrow$~ $x \equiv_V y$ for all $V$ in $\Kst$.
\end{enumerate}
\end{stelling} 

The properties of memorizing and static valuation algebras are stronger than they may appear at first. This is shown by the following proposition, the proof of which can be found in the appendix.

\begin{propositie}\label{memensthandig} 
Let $V$ be a valuation algebra.
\begin{enumerate}
\item If $V$ is memorizing then
\begin{align*}
x/(y \bu x \bu H) &= x/H, &
x \bu y \bu x \bu H &= y \bu x \bu H
\end{align*}
for all $H \in V$ and all formulas $x$, $y$. 
\item If $V$ is static then $x/(y \bu H) = x/H$ for all $H \in V$ and all formulas $x$, $y$.
\end{enumerate}
\end{propositie} 

The property stated in Proposition~\ref{memensthandig}b is especially strong. It says that, no matter what formula $y$ we evaluate, its derivative does not alter the evaluation of a formula $x$. This renders side-effects useless. The following two propositions emphasise this.

\begin{propositie} \label{trivialisst} 
Every trivial valuation algebra is static.
\end{propositie} 

\begin{bewijs}
It is easy to check that a valuation algebra where $a \bu H = H$ for all $a \in \A$ and all valuations $H$, is repetition-proof, contractive, memorizing and static. Clearly all trivial valuation algebras have that property. \qed
\end{bewijs}

\begin{propositie} \label{stistrivial} 
Let $(V, /, \bu)$ be a static valuation algebra and let $H \in V$ be fixed. There exists a trivial valuation algebra $(\{H\}, /_0, \bu_0)$ such that $x/H = x/_0 H$ for every formula $x$.
\end{propositie} 

\begin{bewijs}
Let $(V, /, \bu)$ be a static valuation algebra and let $H \in V$. We construct the valuation algebra $(\{H\}, /_0, \bu_0)$ by stating $a /_0 H = \true$ if and only if $a/H = \true$, and $a \bu_0 H = H$ for all $a \in \A$.

We will prove by induction that $x/H = x/_0 H$ for all formulas $x$. The cases $\true$ and $a$ for $a \in \A$ are clear. Also, if $x = \neg x_1$ is a formula such that $x_1/H = x_1/_0 H$, then $(\neg x_1)/H = \neg (x_1/H) = \neg (x_1/_0H) = (\neg x_1) /_0 H$.
So suppose $x = x_1 \sand x_2$ such that $x_i/H = x_i/_0 H$. Then we use Proposition~\ref{memensthandig} to get
\begin{align*}
x/H
&= \left\{ \begin{array}{ll}
	x_2/(x_1 \bu H) & \text{if~} x_1/H = \true \\
	\false & \text{otherwise}
	\end{array} \right. \\
&= \left\{ \begin{array}{ll}
	x_2/H \phantom{(x_1 \bu~\,)} & \text{if~} x_1/H = \true \\
	\false & \text{otherwise}
	\end{array} \right. \\
&= \left\{ \begin{array}{ll}
	x_2/_0H \phantom{(x_1 \bu\,)} & \text{if~} x_1/_0H = \true \\
	\false & \text{otherwise}
	\end{array} \right. \\
&= x/_0 H.
\end{align*}
This concludes the proof. \qed
\end{bewijs}

This shows that for any valuation $H$ in a static valuation algebra, evaluating a formula in $H$ is essentially the same as evaluating it in a propositional logic sense. However, this proposition does not imply that all static valuation algebras are somehow `equivalent' to trivial valuation algebras. One might imagine a static valuation algebra consisting of multiple valuations, but without any `transitions' between the valuations. A formula evaluated in different valuations of such a valuation algebra could have different outcomes, which is impossible in a trivial valuation algebra.

Still, these two propositions show that $\SSCL$ is arguably the least interesting short-circuit logic in terms of evaluation and satisfiability.

\section{Satisfiability}
\label{sect_satis}

Now that we have defined what it means to evaluate a formula, we can define what it means for a formula to be satisfiable.

\begin{definitie} 
Let $K$ be a collection of valuation algebras. A formula $x$ is \emph{satisfiable} with respect to $K$ if there exists a $V$ in $K$ such that $x/H = \true$ for some $H \in V$, and we denote this by $\SAT_K(x)$. A formula $x$ is \emph{falsifiable} w.r.t. $K$ if there exists a $V$ in $K$ such that $x/H = \false$ for some $H \in V$, and we denote this by $\FAL_K(x)$.
\end{definitie} 

Thus, to show that a formula is satisfiable, it is enough to find or construct a valuation algebra that `satisfies' the formula. Conversely, to show that a formula is not satisfiable, we need to prove that for every valuation in every valuation algebra within a certain collection, the formula evaluates to $\false$. It is not enough to show that the formula is falsifiable; in fact, most formulas will be both satisfiable and falsifiable.
Also, note that $\FAL_K(x) \Leftrightarrow \SAT_K(\neg x)$. This means that for every formula $x$, at least one of $\SAT_K(x)$ and $\FAL_K(x)$ must be true.

Also, if we already have a valuation algebra that satisfies a formula, then it may be part of multiple collections and therefore prove multiple types of satisfiability. In particular, the collections $\Kfr$, $\Krp$, $\Kcn$, $\Kmem$ and $\Kst$ are related, so we immediately find the following proposition.

\begin{propositie} \label{geordend} 
Let $x$ be a formula, then
\[
\SATst(x) \Rightarrow \SATmem(x) \Rightarrow \SATcn(x) \Rightarrow \SATrp(x) \Rightarrow \SATfr(x).
\]
\end{propositie} 

\begin{bewijs}
This follows directly from the definition. \qed
\end{bewijs}

The following theorem further strengthens the connection between the five short-circuit logics and our definition of satisfiability.

\begin{propositie}\label{equivimpsat} 
Let $K$ be a collection of valuation algebras, and let $x$ and $y$ be formulas. If $x \equiv_V y$ for all $V$ in $K$, then $\SAT_K(x) \Leftrightarrow \SAT_K(y)$.
\end{propositie} 

\begin{bewijs}
Let $K$ be a collection of valuation algebras, and let $x$ and $y$ be formulas such that $x \equiv_V y$ for all $V$ in $K$. If $\SAT_K(x)$, then there exists a $V_0$ in $K$ such that $x/H_0 = \true$ for some $H_0 \in V_0$. Because $x \equiv_{V_0} y$, we find $y/H_0 = \true$, and thus $\SAT_K(y)$. If $\neg \SAT_K(x)$, then for every $V$ in $K$, it must be that $x/H = \false$ for all $H \in V$. But for every $V$ in $K$ we have $x \equiv_V y$, thus $y/H = \false$ for all $H \in V$. This shows $\neg \SAT_K(y)$. \qed
\end{bewijs}

\begin{stelling} \label{satcorrect} 
Let $x$ and $y$ be formulas.
\begin{enumerate}
\item If $\EqFSCL \vdash x = y$, then $\SATfr(x) \Leftrightarrow \SATfr(y)$.
\item If $\EqRPSCL \vdash x = y$, then $\SATrp(x) \Leftrightarrow \SATrp(y)$.
\item If $\EqCSCL \vdash x = y$, then $\SATcn(x) \Leftrightarrow \SATcn(y)$.
\item If $\EqMSCL \vdash x = y$, then $\SATmem(x) \Leftrightarrow \SATmem(y)$.
\item If $\EqSSCL \vdash x = y$, then $\SATst(x) \Leftrightarrow \SATst(y)$.
\end{enumerate}
\end{stelling} 

\begin{bewijs}
This follows by combining Theorem~\ref{equivissound} and Proposition~\ref{equivimpsat}. \qed
\end{bewijs}

Lastly, the following corollary reinforces the idea that $\SSCL$ and propositional logic are very similar, especially regarding satisfiability.

\begin{gevolg} \label{satssclisbinsat} 
Let $x$ be a formula. Then $\SATst(x)$ if and only if one can assign either `true' or `false' to each $a \in \A$ such that $x$, as a propositional formula, is true.
\end{gevolg} 

\begin{bewijs}
This follows from Proposition~\ref{trivialisst} and Proposition~\ref{stistrivial}. \qed
\end{bewijs}

\chapter{Path-Satisfiability}
\label{ch_pathsat}

The definitions of evaluation and satisfiability discussed in the previous chapter match the theoretical desires we have for them. However, implementing them seems impossible, or at least highly impractical. They allow all kinds of valuations, which is good, but this generic and abstract nature does not fit the finite and discrete world of a computer program. We therefore need to define an alternative form of evaluation.

As we have already seen how evaluation trees emulate the short-circuit behaviour of our formulas, we will use them as a basis. The basic idea is that a formula can be made true if there is a route, or a `path', through its $\se$-tree to a $\seT$ leaf. We will formalise this by defining `valuation paths' and their result on trees.

\section{Valuation Paths}
\label{sect_valpath}

\begin{definitie} 
A \emph{valuation path} of length $n$ is a sequence $\langle p_1, \dots, p_n \rangle$, where each $p_i$ is a pair $(u_i, b_i) \in \A \times \B$.
\end{definitie} 

Each of the segments of a valuation path consists of an atom from $\A$ and a truth value that states whether this atom should be true or not. There is one valuation path of length $0$, which we will call $\e$.
If $P$ is a valuation path of length $n$, we write $|P| = n$.
To effectively work with valuation paths, we need to be able to manipulate them by adding other valuation paths to them.

\begin{definitie} 
Let $P = \langle p_1, \dots, p_n \rangle$ and $Q = \langle q_1, \dots, q_m \rangle$ be two valuation paths of length $n$ and $m$ respectively. The \emph{concatenation} of $P$ and $Q$ is the valuation path $P \concat Q := \langle p_1, \dots, p_n, q_1, \dots, q_m \rangle$ of length $n+m$.
\end{definitie} 

Note that concatenating $\e$ to a valuation path has no effect, that is, $\e \concat P = P = P \concat \e$. We will also want to use induction and recursion on valuation paths; to this end, note that every valuation path $P$ of positive length can be made by concatenating its first segment with the rest of the valuation path. Thus $P = (u, b) \concat Q$ for some $u \in \A$, some $b \in \B$ and some valuation path $Q$ with $|P| = |Q| + 1$.

Using this, we can now define a valuation path's `result' on a tree. If we apply a valuation path starting with an atom $u \in \A$ to a tree with the same atom $u$ as its root, then the truth value $b$ associated with it determines whether we proceed with the left or the right branch. We iterate this process, until we reach a leaf. If it is a $\seT$ leaf, the result is $\true$, and if it is a $\seF$ leaf, the result is $\false$. However, we must also consider the cases where the valuation path and the tree do not match up. In these cases, we leave the the result undefined. Figure~\ref{variousresults} shows this. Formally:

\begin{definitie} 
The \emph{result} of a valuation path $P$ on a tree $X$, denoted $P \tr X$, is either an element of $\B$ or $\error$. We define $P \tr X$ recursively, as follows:
\begin{align*}
\e \tr \seT &= \true \\
\e \tr \seF &= \false \\
((u, b) \concat Q) \tr (X_1 \sel a \ser X_2) &= \left\{ \begin{array}{ll}
	Q \tr X_1 & \text{if~} u = a \text{~and~} b = \true \\
	Q \tr X_2 & \text{if~} u = a \text{~and~} b = \false
	\end{array} \right.
\end{align*}
and for all other circumstances, we leave $P \tr X$ undefined.
\end{definitie} 

\begin{figure}
\centering
\subfigure[A valid result.]{%
\tikzsetnextfilename{pathresulta}
\begin{tikzpicture}[->,shorten >=1pt,auto,node distance=1.5cm,
  thick,main node/.style={fill=green!20,circle, inner sep=0pt, minimum size=20pt,draw,font=\sffamily\bfseries},
  end node/.style={fill=blue!20,draw,font=\sffamily\bfseries}]

	\node[main node, draw=red!90] (1) {$a$};
	\node[end node] (1r) [below right of=1] {$\false$};
	\path[->] (1) edge (1r);
	\node[main node, draw=red!90] (1e) [below left of=1] {$b$};
	\path[->, draw=red!90] (1) edge (1e);
	\node[end node] (1ee) [below left of=1e] {$\false$};
	\path[->] (1e) edge (1ee);
	\node[end node, fill=red!20, draw=red!90] (1er) [below right of=1e] {$\true$};
	\path[->, draw=red!90] (1e) edge (1er);
\end{tikzpicture}
}%
\hspace{10mm}
\subfigure[Invalid atom.]{%
\tikzsetnextfilename{pathresultb}
\begin{tikzpicture}[->,shorten >=1pt,auto,node distance=1.5cm,
  thick,main node/.style={fill=green!20,circle, inner sep=0pt, minimum size=20pt,draw,font=\sffamily\bfseries},
  end node/.style={fill=blue!20,draw,font=\sffamily\bfseries}]

	\node[main node, draw=red!90] (1) {$a$};
	\node[end node] (1r) [below right of=1] {$\false$};
	\path[->] (1) edge (1r);
	\node[main node, fill=yellow!20, draw=red!90] (1e) [below left of=1] {$b$};
	\path[->, draw=red!90] (1) edge (1e);
	\node[end node] (1ee) [below left of=1e] {$\false$};
	\path[->] (1e) edge (1ee);
	\node[end node] (1er) [below right of=1e] {$\true$};
	\path[->] (1e) edge (1er);
\end{tikzpicture}
}%
\\
\subfigure[Too short.]{%
\tikzsetnextfilename{pathresultc}
\begin{tikzpicture}[->,shorten >=1pt,auto,node distance=1.5cm,
  thick,main node/.style={fill=green!20,circle, inner sep=0pt, minimum size=20pt,draw,font=\sffamily\bfseries},
  end node/.style={fill=blue!20,draw,font=\sffamily\bfseries}]

	\node[main node, draw=red!90] (1) {$a$};
	\node[end node] (1r) [below right of=1] {$\false$};
	\path[->] (1) edge (1r);
	\node[main node] (1e) [below left of=1] {$b$};
	\path[->, draw=red!90] (1) edge (1e);
	\node[end node] (1ee) [below left of=1e] {$\false$};
	\path[->] (1e) edge (1ee);
	\node[end node] (1er) [below right of=1e] {$\true$};
	\path[->] (1e) edge (1er);
\end{tikzpicture}
}%
\hspace{10mm}
\subfigure[Too long.]{%
\tikzsetnextfilename{pathresultd}
\begin{tikzpicture}[->,shorten >=1pt,auto,node distance=1.5cm,
  thick,main node/.style={fill=green!20,circle, inner sep=0pt, minimum size=20pt,draw,font=\sffamily\bfseries},
  end node/.style={fill=blue!20,draw,font=\sffamily\bfseries}]

	\node[main node, draw=red!90] (1) {$a$};
	\node[end node] (1r) [below right of=1] {$\false$};
	\path[->] (1) edge (1r);
	\node[main node, draw=red!90] (1e) [below left of=1] {$b$};
	\path[->, draw=red!90] (1) edge (1e);
	\node[end node] (1ee) [below left of=1e] {$\false$};
	\path[->] (1e) edge (1ee);
	\node[end node, fill=yellow!20, draw=red!90] (1er) [below right of=1e] {$\true$};
	\path[->, draw=red!90] (1e) edge (1er);
\end{tikzpicture}
}%
\caption{The result of the valuation path $\langle (a, \true), (b, \false) \rangle$ in the tree $se(a \protect \sand \neg b)$ is $\true$.
The results of $\langle (a, \true), (a, \false) \rangle$, $\langle (a, \true) \rangle$ and $\langle (a, \true), (b, \false), (a, \false) \rangle$ in that same tree are all $\error$.}
\label{variousresults}
\end{figure}

Eventually, we want to relate this back to formulas, as generic trees are not the most interesting to us. In the rest of this section, we will discuss what results valuation paths have on $\se$-trees.

First, consider the following: we have two trees, $X$ and $Y$, and a path $P$. If $P$ results to either $\true$ of $\false$ on $X$, then that means that $P$ leads us through $X$ to one of the leaves of $X$. If where to replace this leaf with $Y$, then $P$ would lead us to the root of $Y$. Intuitively, we want to be able to continue the path where we left of and traverse $Y$ as well, by appending another path to $P$. The following proposition allows us to do so.

\begin{propositie} \label{resultconcat}  
Let $X$, $Y$, $Y'$ be trees and $P$, $Q$ paths. If $P \tr X$ is defined,
then $(P \concat Q) \tr X[\beta \mapsto Y, \neg \beta \mapsto Y'] = Q \tr Y$ where $\beta = P \tr X$.
\end{propositie} 

\begin{bewijs}
Let $Y$ and $Y'$ be trees and $Q$ a path. We prove this proposition by induction to the depth of $X$. Call $X$ ``compatible'' if, for all paths $P$, 
\[
\text{if}~\beta = P \tr X ~\text{is defined, then}~ (P \concat Q) \tr X[\beta \mapsto Y, \neg\beta \mapsto Y'] = Q \tr Y.
\]
The only trees of depth $0$ are $\seT$ and $\seF$. Let $X$ be either. If $P$ is a path such that $P \tr X$ is defined, then $P = \e$, thus $P \concat Q = Q$. If $X = \seT$ then $P \tr X = \true$, which means $Z = \seT[\seT \mapsto Y, \seF \mapsto Y']$; if not, then $X = \seF$, $P \tr X = \false$ and $Z = \seF[\seT \mapsto Y', \seF \mapsto Y]$. In either case, $Z = Y$, thus $(P \concat Q) \tr Z = Q \tr Z = Q \tr Y$. We conclude that all trees of depth $0$ are compatible.

Let $n \ge 0$ and assume that all trees of depth at most $n$ are compatible. Let $X$ be a tree of depth $n+1$, then $X = X_1 \sel a \ser X_2$ for trees $X_1$ and $X_2$ and for some $a \in \A$. Then $X_1$ and $X_2$ are of depth at most $n$, thus compatible. To complete the proof, we need to show that $X$ is compatible.

Let $P$ be a path such that $\beta = P \tr X$ is defined, then $P$ must be of the form $P = (a, b) \concat R$ for some $b \in \B$ and some path $R$. 
We get
\begin{align*}
\beta = P \tr X = ((a, b) \concat R) \tr (X_1 \sel a \ser X_2) &= \left\{ \begin{array}{ll}
	R \tr X_1 & \text{if~} b = \true \\
	R \tr X_2 & \text{if~} b = \false \\
	\end{array} \right.
\end{align*}
and this means that if $b = \true$ then $R \tr X_1 = \beta$, and if $b = \false$ then $R \tr X_2 = \beta$.
Let $Z = X[\beta \mapsto Y, \neg \beta \mapsto Y']$, then $Z = Z_1 \sel a \ser Z_2$ where $Z_i = X_i[\beta \mapsto Y, \neg \beta \mapsto Y']$. 
Because $X_1$ and $X_2$ are compatible, we find
\begin{align*}
(P \concat Q) \tr Z
&= ((a, b) \concat (R \concat Q)) \tr (Z_1 \sel a \ser Z_2) \\
&= \left\{ \begin{array}{ll}
	(R \concat Q) \tr Z_1 & \text{if~} b = \true \\
	(R \concat Q) \tr Z_2 & \text{if~} b = \false \\
	\end{array} \right. \\
&= \left\{ \begin{array}{ll}
	Q \tr Y & \text{if~} b = \true \\
	Q \tr Y & \text{if~} b = \false \\
	\end{array} \right. \\
&= Q \tr Y.
\end{align*}
Therefore $X$ is compatible. \qed
\end{bewijs}

The next proposition allows us to say something useful about the results of valuation paths on $\se$-trees; namely that they are what we might expect them to be.

\begin{propositie} \label{resultlogic}  
Let $x$, $y$ be formulas and $P$, $Q$ paths. If $P \tr \se(x)$ is defined, then
\begin{align*}
P \tr \se(\neg x) &= \neg (P \tr \se(x)) \\
(P \cdot Q) \tr \se(x \sand y) &= \left\{ \begin{array}{ll}
	Q \tr \se(y) & \text{if~} P \tr \se(x) = \true \\
	Q \tr \seF & \text{otherwise} \\
	\end{array} \right.
\end{align*}
\end{propositie} 

\begin{bewijs}
Let $x$ be a formula, let $X = \se(x)$ and let $P$ a path such that $P \tr X$ is defined. Let $Q = \e$ so that $P \concat Q = P$. If $P \tr X = \true$, then let $Y = \seF$ and $Y' = \seT$ which gives us $\se(\neg x) = X[\seT \mapsto Y, \seF \mapsto Y']$. Now we can use Proposition~\ref{resultconcat} in order to get $P \tr \se(\neg x) = Q \tr Y = \e \tr \seF = \false$. Otherwise let $Y = \seT$ and $Y' = \seF$, which gives $\se(\neg x) = X[\seT \mapsto Y', \seF \mapsto Y]$. By the proposition, $P \tr \se(\neg x) = Q \tr Y = \e \tr \seT = \true$. Either way, we find $\neg(P \tr \se(x))$.

Let $x, y$ be formulas, $X = \se(x)$ and let $P, Q$ paths such that $P \tr X$ is defined. If $P \tr X = \true$, then let $Y = \se(y)$ and $Y' = \seF$, thus $\se(x \sand y) = X[\seT \mapsto Y, \seF \mapsto Y']$. The proposition tells us $(P \concat Q) \tr \se(x \sand y) = Q \tr \se(y)$. Otherwise, let $Y = \seF$ and $Y' = \se(y)$, and thus $\se(x \sand y) = X[\seT \mapsto Y', \seF \mapsto Y]$. Thus $(P \concat Q) \tr \se(x \sand y) = Q \tr \seF$ by the proposition. \qed
\end{bewijs}

We also want a converse to the previous proposition. That is, if a path traverses a `compound' tree to a leaf of that tree, then some initial part of this path will lead us to the point where the substitution took place. More formally:

\begin{propositie} \label{resultdivis} 
Let $X$, $Y$, $Y'$ be trees and $P$ a path. If $P : X[\seT \mapsto Y, \seF \mapsto Y']$ is defined, then there are paths $R$ and $Q$ with $P = R \concat Q$, such that $R \tr X$ is defined and
\begin{align} \label{divisifif} \tag{$\star$}
P \tr X[\seT \mapsto Y, \seF \mapsto Y'] &= \left\{ \begin{array}{ll}
	Q \tr Y & \text{if~} R \tr X = \true \\
	Q \tr Y' & \text{otherwise} \\
	\end{array} \right.
\end{align}
\end{propositie} 

\begin{bewijs}
Let $Y$ and $Y'$ be trees. We also prove this proposition by induction, but this time to the length of $P$. Call $P$ ``divisible'' if for every tree $X$ there are $R, Q$ such that $P = R \concat Q$ and 
\[
\text{if~} P \tr X[\seT \mapsto Y, \seF \mapsto Y'] \text{~is defined, then~} R \tr X \text{~is defined and (\ref{divisifif})}.
\]

The only path of length $P$ is $\e$. Let $X$ be a tree and let $Z = X[\seT \mapsto Y, \seF \mapsto Y']$. Let $R = \e$ and $Q = \e$. If $\e \tr Z$ is defined, then $Z \in \{\seT, \seF\}$, thus $X, Y, Y' \in \{\seT, \seF\}$. If $X = \seT$, then $R \tr X = \true$ and $Z = Y$. If $X = \seF$, then $R \tr X = \false$ and $Z = Y'$. Either way, (\ref{divisifif}) holds and $\e$ is divisible.

Let $n \ge 0$ and assume all paths of length at most $n$ are divisible. Let $P$ be of length $n+1$. To complete the proof, we need to show that $P$ is divisible. Note that if $X \in \{\seT, \seF\}$ and $Z = X[\seT \mapsto Y, \seF \mapsto Y']$, then either $Z = Y$ or $Z = Y'$ and we can take $R = \e$ and $Q = P$ to immediately get the result. Thus in the following we assume that $X = X_1 \sel a \ser X_2$ for some trees $X_1$, $X_2$ and some $a \in \A$, and this gives us $Z = Z_1 \sel a \ser Z_2$ where $Z_i = X_i[\seT \mapsto Y, \seF \mapsto Y']$.

Because $P \ne \e$, we can write $P = (u, b) \concat P'$ for some $u \in \A$, some $b \in \B$ and some path $P'$ of length $n$. Suppose $P \tr Z$ is defined, then $u = a$ and we get
\begin{align*}
P \tr Z = ((a, b) \concat P') \tr (Z_1 \sel a \ser Z_2) 
&= \left\{ \begin{array}{ll}
	P' \tr Z_1 & \text{if~} b = \true \\
	P' \tr Z_2 & \text{otherwise} \\
	\end{array} \right.
\end{align*}
thus $P' \tr Z_i = P \tr Z$ is defined, where $i = 1$ if $b = \true$ and $i = 2$ otherwise.

Because $P'$ is divisible, there are $R', Q'$ such that $P' = R', Q'$, that $R' \tr X_i$ is defined and 
\begin{align*}
P' \tr Z_i &= \left\{ \begin{array}{ll}
	Q' \tr Y & \text{if~} R' \tr X_i = \true \\
	Q' \tr Y' & \text{otherwise} \\
	\end{array} \right.
\end{align*}
Take $R = (u, b) \concat R'$ and $Q = Q'$, then $P = (u,b) \concat P' = ((u, b) \concat R') \concat Q'$. We find that $R \tr X = R' \tr X_i$ is defined by our choice of $i$, and (\ref{divisifif}) follows. Thus $P$ is divisible. \qed
\end{bewijs}

\section{Path-Satisfiability}
\label{sect_pathsat}

In the previous section, we have defined an alternative way to evaluate formulas, based on their $\se$-tree. Using this, we can now define our alternative satisfiability, called `path-satisfiability'. In principle, a formula is path-satisfiable if there is a path that results in $\true$ on the formula's $\se$-tree, and path-falsiable if there is a path that results in $\false$.

However, this definition alone gives us no method allow or disallow certain side-effects, which we need to correspond to our five short-circuit logics. To this purpose we define two properties for valuation paths: `repetition-proof' and `memorizing'.

\begin{definitie} 
Let $P = \langle (u_1, b_1), \dots, (u_n, b_n) \rangle$ be a valuation path, then $P$ is called
\begin{itemize}
\item \emph{repetition-proof} if $u_i = u_{i+1} \Longrightarrow b_i = b_{i+1}$ for all $i < n$.
\item \emph{memorizing} if $u_i = u_j \Longrightarrow b_i = b_j$ for all $i, j \le n$.
\end{itemize}
\end{definitie} 

Of course, every memorizing valuation path is also repetition-proof. Now we can formally define three forms of path-satisfiability; one `free' path-satisfiability that is without any requirements, and one path-satisfiability for each of the two properties defined above.

\begin{definitie} 
Let $x$ be a formula. A formula is \emph{path-satisfiable} if there exists a valuation path $P$ such that $P \tr \se(x) = \true$, and we denote this $\PATHSATfr(x)$. A formula is \emph{rp-path-satisfiable}, denoted $\PATHSATrp(x)$, if there is a repetition-proof path, and \emph{mem-path-satisfiable}, denoted $\PATHSATmem(x)$, if there is a memorizing path. 

We also define three analogous forms of path-falsifiability, where $P \tr \se(x) = \false$, and we denote these by $\PATHFALfr(x)$, $\PATHFALrp(x)$ and $\PATHFALmem(x)$.
\end{definitie} 

If a tree has no $\seT$ leaves, then there clearly cannot be a valuation path that results in $\true$ on this tree. It is not hard to see that if all kinds of valuation path are allowed, the converse is also true; if a tree has a $\seT$ leaf, then there is a valuation path that results in $\true$ on this tree. This is stated by the following proposition.

\begin{propositie} \label{pathsatleaves} 
Let $x$ be a formula.
\begin{enumerate}
\item $\PATHSATfr(x)$ if and only if $\se(x)$ has a $\seT$ leaf.
\item $\PATHFALfr(x)$ if and only if $\se(x)$ has a $\seF$ leaf.
\end{enumerate}
\end{propositie} 

\begin{bewijs}
If a tree of the form $X \sel a \ser Y$ contains a leaf, then this leaf can be reached by a valuation path either of the form $(a, \true) \concat P$ where $P$ runs through $X$, or of the form $(a, \false) \concat Q$ where $Q$ runs through $Y$. From this, both statements follow. \qed
\end{bewijs}

This proposition has two corollaries that relate to constant-free formulas and formulas in normal form.

\begin{gevolg} \label{opentreesat2} 
If $x$ is constant-free formula, then $\PATHSATfr(x)$ and $\PATHFALfr(x)$.
\end{gevolg} 

\begin{bewijs}
This follows from Corollary~\ref{constantfreeopen} and Proposition~\ref{pathsatleaves}. \qed
\end{bewijs}

\begin{gevolg} \label{normalformpathsat} 
Let $x$ be a formula. 
\begin{enumerate}
\item $\PATHSATfr(x)$ and $\neg \PATHFALfr(x)$ if and only if $\snf(x)$ is a $\true$-term.
\item $\neg \PATHSATfr(x)$ and $\PATHFALfr(x)$ if and only if $\snf(x)$ is a $\false$-term.
\item $\PATHSATfr(x)$ and $\PATHFALfr(x)$ if and only if $\snf(x)$ is a $\true*$-term.
\end{enumerate}
\end{gevolg} 

\begin{bewijs}
This follows by combining Proposition~\ref{normalformfundamental} and Proposition~\ref{pathsatleaves}. \qed
\end{bewijs}

For repetition-proof and memorizing paths, a weaker version of this last corollary exists.

\begin{gevolg} \label{normalformpathsat2} 
Let $x$ be a formula.
\begin{enumerate}
\item If $\snf(x)$ is a $\true$-term, then $\PATHSATmem(x)$ and $\neg \PATHFALmem(x)$.
\item If $\snf(x)$ is a $\false$-term, then $\neg \PATHSATmem(x)$ and $\PATHFALmem(x)$.
\end{enumerate}
\end{gevolg} 

\begin{bewijs}
We can certainly construct a memorizing valuation path $P$ such that $P \tr \se(x)$ is defined, for example by simply assigning $\true$ to all atoms. By Proposition~\ref{normalformfundamental}, if $\snf(x)$ is a $\true$-term then $\se(x)$ is closed by $\true$. This means $P \tr \se(x) = \true$. And of course, if $\se(x)$ has no $\false$-leaves, then no valuation path $Q$ exists with $Q \tr \se(x) = \false$. Analogous statements can be made when $\snf(x)$ is a $\false$-term. \qed
\end{bewijs}

These three corollaries may suggest that path-satisfiability is somewhat trivial to solve. However, most formulas will not be constant-free, and in Chapter~\ref{ch_implem} we will discuss how normal forms are not ideal to solve path-satisfiability.

Before we continue, an analogue to Proposition~\ref{geordend}.

\begin{propositie} \label{pathgeordend} 
Let $x$ be a formula, then
\[
\PATHSATmem(x) \Longrightarrow \PATHSATrp(x) \Longrightarrow \PATHSATfr(x)
\]
\end{propositie} 

\begin{bewijs}
This follows directly from the definitions. \qed
\end{bewijs}

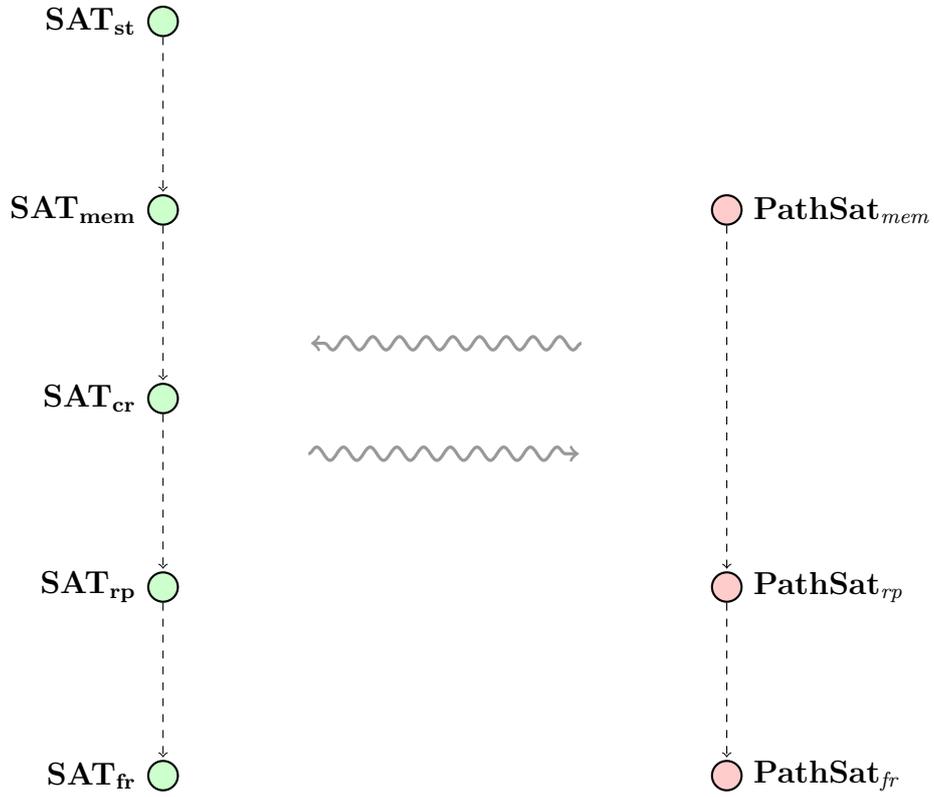
\begin{figure}
\centering
\tikzsetnextfilename{prachtig1}
\begin{tikzpicture}[->,shorten >=1pt,auto,node distance=25mm,
  main node/.style={thick, circle,fill=green!20,draw, minimum size=10pt}]

  \node[main node] (1) [label=left:{$\SATfr$}] {};
  \node[main node] (2) [label=left:{$\SATrp$}, above of=1] {};
  \node[main node] (3) [label=left:{$\SATcn$}, above of=2] {};
  \node[main node] (4) [label=left:{$\SATmem$}, above of=3] {};
  \node[main node] (5) [label=left:{$\SATst$}, above of=4] {};

  \path[dashed, every node/.style={font=\sffamily\small}]
    (5) edge (4)
    (4) edge (3)
    (3) edge (2)
    (2) edge (1)
	;

  \node[main node, fill=red!20] (11) [label=right:{$\PATHSATfr$}, right=70mm of 1] {};
  \node[main node, fill=red!20] (22) [label=right:{$\PATHSATrp$}, right=70mm of 2] {};
  \node[main node, fill=red!20] (44) [label=right:{$\PATHSATmem$}, right=70mm of 4] {};

  \path[dashed, every node/.style={font=\sffamily\small}]
    (44) edge (22)
    (22) edge (11)
	;

  \node (hmm1) [above right of=2] {};
  \node (hmm2) [below right of=4] {};
  \node (hmm3) [above left of=22] {};
  \node (hmm4) [below left of=44] {};

  \path[very thick, draw=black!40]
	(hmm1) edge[decorate, decoration={snake, post length=1mm}] (hmm3)
	(hmm4) edge[decorate, decoration={snake, post length=1mm}] (hmm2)
	;
\end{tikzpicture}
\caption[A first overview of satisfiability and path-satisfiability.]{A schematic overview of satisfiability and path-satisfiability.
The five green nodes on the left represent satisfiability, as described in Section~\ref{sect_satis} for the five logics described in Section~\ref{sect_scl}. The descending dashed arrows between them are given by Proposition~\ref{geordend}.
The three red nodes on the right represent the three types of path-satisfiability defined in Section~\ref{sect_pathsat}, and the descending arrows between them are given by Proposition~\ref{pathgeordend}.
}
\label{prachtig1}
\end{figure}

In Chapter~\ref{ch_implem} we will discuss an implementation of path-satisfiability. However, our original goal was to describe and implement ``real'' satisfiability. If our two forms of evaluation and satisfiability do not match up, we have effectively wasted our time defining and proving something unrelated.
Figure~\ref{prachtig1} illustrates this disconnect. As we will prove the connections between the types of satisfiability, we will update this illustration.

To show a first connection between valuation algebras and valuation paths, consider the following: suppose we have a formula $x$ that we are evaluating in some valuation algebra, and suppose we make a note each time we encounter an atom, both of which atom it is and of what truth value it is assigned. Then at the end we have a `diary' of sorts, and this diary is in fact a valuation path. The following definition formalises this procedure.

\begin{definitie} 
Let $V$ be a valuation algebra. For a formula $x$ and a valuation $H \in V$, we define the \emph{evaluation path} of $x$ at $H$, denoted by $x \print H$, as follows:
\begin{align*}
\true \print H &= \e \\
a \print H &= \langle (a, a/H) \rangle \\
(\neg x) \print H &= x \print H \\
(x \sand y) \print H &= \left\{ \begin{array}{ll}
	(x \print H) \concat (y \print (x \bu H)) & \text{if~} x/H = \true \\
	\phantom{(}x \print H & \text{otherwise}
	\end{array} \right.
\end{align*}
\end{definitie} 

The name `evaluation path' refers to how this valuation path is created while evaluating the formula. 
The purpose of these evaluation paths is that the result of an evaluation path on the $\se$-tree of a formula is exactly the same as the evaluation of the formula in the valuation. 
Proposition~\ref{printresult} states this useful fact.

\begin{propositie} \label{printmeaning} 
Let $V$ be a valuation algebra. For a formula $x$ and some $H \in V$, let $x \print H = \langle p_1, \dots, p_n \rangle$ with $p_i = (u_i, b_i)$. Then $b_i = u_i / (u_{i-1} \bu \ldots \bu u_{1} \bu H)$ for $1 \le i \le n$.
\end{propositie} 

\begin{bewijs}
This is easy to check using Proposition~\ref{resultlogic} and Proposition~\ref{resultdivis}.
\qed
\end{bewijs}

\begin{propositie} \label{printresult} 
Let $V$ be a valuation algebra. Then $(x \print H) \tr \se(x) = x/H$ for every formula $x$ and every $H \in V$.
\end{propositie} 

\begin{bewijs}
This is easy to check using Proposition~\ref{printmeaning}.
\qed
\end{bewijs}

This means that for every formula, if there is a valuation algebra where the formula evaluates to $\true$, then there is also a valuation path whose result in the formula's $\se$-tree is $\true$. In fact, this valuation path will have the similar properties to the valuation algebra.

\begin{propositie} \label{printisrp} 
Let $V$ be a valuation algebra, let $x$ be a formula and let $H \in V$. 
\begin{enumerate}
\item If $V$ is repetition-proof, then $x \print H$ is repetition-proof. 
\item If $V$ is memorizing, then $x \print H$ is memorizing.
\end{enumerate}
\end{propositie} 

\begin{bewijs}
This is easy to check using Proposition~\ref{printmeaning}.
\qed
\end{bewijs}

We can now state the following result, which establish one half of the connection between satisfiability and path-satisfiability that we are trying to prove.

\begin{stelling}\label{satispath} 
Let $x$ be a formula.
\begin{enumerate}
\item If $\SATfr(x)$, then $\PATHSATfr(x)$.
\item If $\SATrp(x)$, then $\PATHSATrp(x)$.
\item If $\SATmem(x)$, then $\PATHSATmem(x)$.
\end{enumerate}
\end{stelling} 

\begin{bewijs}
Let $x$ be a formula such that $\SATfr(x)$, and let $V$ be a valuation algebra with $H \in V$ such that $x/H = \true$. Let $P = x \print H$. By Proposition~\ref{printresult}, we have $P \tr \se(x) = \true$. This means $\PATHSATfr(x)$.

If $\SATrp(x)$ (resp. $\SATmem(x)$), then we can find $V$ so that additionally $V$ is repetition-proof (resp. memorizing). By Proposition~\ref{printisrp}, $P$ is repetition-proof (resp. memorizing), which means $\PATHSATrp(x)$ (resp. $\PATHSATmem(x)$). \qed
\end{bewijs}

Now we have shown an important connection. Figure~\ref{prachtig2} illustrates this. The next two sections will be spent establishing a converse connection.

\begin{figure}
\centering
\tikzsetnextfilename{prachtig2}
\begin{tikzpicture}[->,shorten >=1pt,auto,node distance=25mm,
  main node/.style={thick, circle,fill=green!20,draw, minimum size=10pt}]

  \node[main node] (1) [label=left:{$\SATfr$}] {};
  \node[main node] (2) [label=left:{$\SATrp$}, above of=1] {};
  \node[main node] (3) [label=left:{$\SATcn$}, above of=2] {};
  \node[main node] (4) [label=left:{$\SATmem$}, above of=3] {};
  \node[main node] (5) [label=left:{$\SATst$}, above of=4] {};

  \path[dashed, every node/.style={font=\sffamily\small}]
    (5) edge (4)
    (4) edge (3)
    (3) edge (2)
    (2) edge (1)
	;

  \node[main node, fill=red!20] (11) [label=right:{$\PATHSATfr$}, right=70mm of 1] {};
  \node[main node, fill=red!20] (22) [label=right:{$\PATHSATrp$}, right=70mm of 2] {};
  \node[main node, fill=red!20] (44) [label=right:{$\PATHSATmem$}, right=70mm of 4] {};

  \path[dashed, every node/.style={font=\sffamily\small}]
    (44) edge (22)
    (22) edge (11)
	;

  \path[very thick, every node/.style={font=\sffamily\small}]
    (1) edge node[below] {$\print$} (11)
    (2) edge node[below] {$\print$} (22)
    (4) edge node[below] {$\print$} (44)
	;
\end{tikzpicture}
\caption[A second overview of satisfiability and path-satisfiability.]{An updated overview, based on Figure~\ref{prachtig1}.
The three thick arrows labeled $\print$ are given by Theorem~\ref{satispath}.
}
\label{prachtig2}
\end{figure}

\section{Norm-based Constructors}
\label{sect_constru}

To show a connection between path-satisfiability and satisfiability, we need to solve the following problem: suppose we have found a valuation path that results in $\true$ on the $\se$-tree of a given formula; how do we create a valuation algebra where the formula evaluates to $\true$?
At first glance, this seems relatively easy since we can add as many valuations as we need, and each valuation can assign whichever truth value we want to each atom. For each atom we come across, we make a valuation that makes this atom true and then we jump to the next valuation for the next atom. Thus, for a path $P = \langle (u_1, b_1), \dots, (u_n, b_n) \rangle$ we might make a valuation algebra $(\N, /, \bu)$ such that $a/i = b_i$ and $a \bu i = i+1$ for all $i$, as depicted in Figure~\ref{bijnavapic}.

\begin{figure}
\centering
\tikzsetnextfilename{bijnavapic}
\begin{tikzpicture}[->,shorten >=1pt,auto,node distance=17mm, thick]
\tikzstyle{main node}=[circle,fill=orange!20,draw,font=\sffamily\bfseries]
\tikzstyle{end node}=[circle,fill=gray!10,draw,font=\sffamily\bfseries]

  \node[main node] (1) {1};
  \node[main node] (2) [right of=1] {2};
  \node[main node] (3) [right of=2] {3};
  \node[main node] (4) [right of=3] {4};
  \node[main node] (5) [right of=4] {5};
  \node[main node] (6) [right of=5] {6};
  \node (7) [right of=6] {...};

  \path[every node/.style={font=\sffamily\small}]
    (1) edge node[above] {a.T} (2)
    (2) edge node[above] {b.F} (3)
    (3) edge node[above] {b.F} (4)
    (4) edge node[above] {b.T} (5)
    (5) edge node[above] {a.F} (6)
    (6) edge node[above] {a.F} (7);
\end{tikzpicture}
\caption{A first attempt to create a valuation algebra $(\N, /, \bu)$ for the valuation path $P = \langle (a, \true), (b, \false), (b, \false), (b, \true), (a, \false), (a, \false) \rangle$.}
\label{bijnavapic}
\end{figure}

Such a valuation algebra could work if the remaining gaps in its definition are filled; however, it has proven difficult to properly write down and prove the propositions that we would need to use such a valuation algebra. We would much rather use a recursive definition, which would allow us to prove our proofs using induction.
Therefore, we will only construct finite valuation algebras, and their size will depend on the ``size'' of the valuation path. We might need different ways to assign a size to a valuation path, and this is achieved by defining norms.

\begin{definitie} 
A \emph{norm} on valuation paths is a function $|| \cdot ||$ that maps a valuation path $P$ to a value $||P|| \ge 0$ such that $|| \e || = 0$ and $|| P \concat Q || \le || P || + || Q ||$.
\end{definitie} 

One norm was already defined in Section~\ref{sect_valpath}: the length norm $| \cdot |$. Note that, for paths $P$ and $Q$, $|P \concat Q| = |P| + |Q|$ and that if $|P| = 0$ then $P = \e$. Traditionally this last property is an additional condition of norms, and functions without it are called ``semi-norms''; however, we ignore this distinction. Therefore, the trivial norm defined by $||P|| = 0$ for all paths $P$ is also a norm.

In this section we will define a few `constructors' that assign a valuation algebra to each valuation path. To effectively use recursion and induction, we need that if a valuation path $P$ is a concatenation of $Q$ and $R$, then the valuation algebra associated with $P$ should somehow resemble a combination of the two valuation algebras associated to $Q$ and $R$. 
However, we do not have a way to combine valuation algebras. Instead, we will try to create constructors that are `invariant' to concatenation; that is, the valuation algebra of a path $P$ is `embedded' in the valuation algebra of any path of the form $R_1 \concat P \concat R_2$. 
These vague notions will be properly defined later in this section. First, we define what kind of constructors we will make.

\begin{definitie} 
Let $|| \cdot ||$ be a norm.
If for each valuation path $P$ a valuation algebra $\constru(P)$ of the form $(\{1, \dots, ||P|| + 1\}, /, \bu)$ is defined such that $i \le (a \bu i) \le i+1$ for all $i$ and all $a \in \A$, then $\constru$ is a \emph{norm-based constructor} for $|| \cdot ||$.
\end{definitie} 

As desired, if $\constru$ is a norm-based constructor then the size of $\constru(P)$ depends linearly on $||P||$. Note that the second property states that for each valuation $i$ and each $a \in \A$, either $a$ does not change $i$ or it advances $i$ by one; it cannot `jump' forward and it cannot go back. This will help us in making these constructors `invariant'.

We are now ready to create our first norm-based constructor, called $\va$.

\begin{definitie}\label{defva} 
Let $P = \langle (u_1, b_1), \dots, (u_n, b_n) \rangle$ be a valuation path. For $a \in \A$ and $k \le n+1$, we define $\last(a, k)$ as the largest $i \le n$ such that $i \le k$ and $u_i = a$, or $0$ if no such $i$ exists. We define $\va(P)$ as the valuation algebra $(\{1, \dots, n+1\}, /, \bu)$, where $/$ and $\bu$ are defined by
\begin{align*}
a / i &= \left\{ \begin{array}{ll}
	b_j ~~~\,& \text{if~} j = \last(a, i) > 0 \\
	\false & \text{otherwise} \\
	\end{array} \right. \\
a \bu i &= \left\{ \begin{array}{ll}
	i+1 & \text{if~} i \le n \text{~and~} u_i = a \\
	i & \text{otherwise} \\
 	\end{array} \right.
\end{align*}
for $a \in \A$ and $i \le n+1$.
\end{definitie} 

Note that $\va$ is a norm-based constructor for the length norm $| \cdot |$. Similar to our earlier idea, the valuation algebra $\va(P)$ for a path $P = \langle (u_1, b_1), \dots, (u_n, b_n) \rangle$ has the desirable properties that $u_i/i = b_i$ and $u_i \bu i = i+1$, but this time for a finite amount of valuations instead of for all $\N$.
As a counterpart to Figure~\ref{bijnavapic}, the valuation algebra $\va(P)$ is depicted in Figure~\ref{voorbeeldvapic} for the same valuation path $P$.

\begin{figure}
\centering
\tikzsetnextfilename{voorbeeldvapic}
\begin{tikzpicture}[->,shorten >=1pt,auto,node distance=17mm, thick]
\tikzstyle{main node}=[circle,fill=orange!20,draw,font=\sffamily\bfseries]
\tikzstyle{end node}=[circle,fill=gray!10,draw,font=\sffamily\bfseries]

  \node[main node] (1) {1};
  \node[main node] (2) [right of=1] {2};
  \node[main node] (3) [right of=2] {3};
  \node[main node] (4) [right of=3] {4};
  \node[main node] (5) [right of=4] {5};
  \node[main node] (6) [right of=5] {6};
  \node[end node] (7) [right of=6] {7};

  \path[every node/.style={font=\sffamily\small}]
    (1) edge node[above] {a.T} (2)
    (2) edge node[above] {b.F} (3)
    (3) edge node[above] {b.F} (4)
    (4) edge node[above] {b.T} (5)
    (5) edge node[above] {a.F} (6)
    (6) edge node[above] {a.F} (7)
    (7) edge[loop right] node[right] {b.T} (7)
    (1) edge[loop below] node[below] {b.F} (1)
    (2) edge[loop below] node[below] {a.T} (2)
    (3) edge[loop below] node[below] {a.T} (3)
    (4) edge[loop below] node[below] {a.T} (4)
    (5) edge[loop below] node[below] {b.T} (5)
    (6) edge[loop below] node[below] {b.T} (6)
    (7) edge[loop below] node[below] {a.F} (7);
\end{tikzpicture}
\caption{An illustration of the valuation algebra $\va(P)$, again for the valuation path $P = \langle (a, \true), (b, \false), (b, \false), (b, \true), (a, \false), (a, \false) \rangle$.}
\label{voorbeeldvapic}
\end{figure}

From our illustrated example, it is clear that $\va(P)$ shares some features with $P$. In Section~\ref{sect_pathissat}, we will prove a very strong result about the norm-based constructor $\va$:

\begin{lemma*}[\ref{pathissatlemma}a]
Let $x$ be a formula and let $P$ a valuation path such that $P \tr \se(x)$ is defined.
Then $x/1 = P \tr \se(x)$ in $\va(P)$.
\end{lemma*}

In particular, if $P \tr \se(x) = \true$, then $x/1 = \true$ in $\va(P)$. As a consequence, each path-satisfiable formula is satisfiable with respect to $\Kfr$, and with the use of the following proposition, each rp-path-satisfiable formula is satisfiable with respect to $\Krp$.

\begin{propositie} \label{varpisrp} 
If $P$ is a repetition-proof valuation path, then $\va(P)$ is a repetition-proof valuation algebra.
\end{propositie} 

\begin{bewijs}
Let $P$ be a repetition-proof valuation path of length $n$. Let $i \le n+1$ and $a \in \A$, then we need to show that $a/(a \bu i) = a/i$ in $\va(P)$. If $a \bu i = i$, then this is clear, so we can suppose that $i \le n$ and $a \bu i = i+1$. This means $u_i = a$ and $a/i = b_i$. Now consider $a/(i+1)$; clearly $i \le \last(a, i+1) \le i+1$, but this means that either $a/(i+1) = b_i$ or $u_{i+1} = a$ and $a/(i+1) = b_{i+1}$. In the latter case, $b_{i+1} = b_i$ follows as $P$ is repetition-proof. \qed
\end{bewijs}

This sounds like a great result, and we can expand on Figure~\ref{prachtig2}.
However, some care must be taken here. If we were to only show a connection from $\PATHSATrp$ to $\SATrp$, and one from $\PATHSATmem$ to $\SATmem$, then we would leave $\SATcn$ and $\SATst$ without path-related equivalents.
This would imply that our three path-satisfiabilities are insufficient to describe the five different satisfiabilities.
Instead, we will show connections from $\PATHSATrp$ directly to $\SATcn$ and similarly from $\PATHSATmem$ to $\SATst$. The implications of this will be discussed in Chapter~\ref{ch_conclu};
for now, we are concerned with constructing appropriate valuation algebras. 

Unfortunately, our example in Figure~\ref{voorbeeldvapic} suggests that the valuation algebras created by $\va$ will not be contractive for most repetition-proof valuation paths, so $\va$ will not do. The problem lies in the following: if a valuation path $P$ has two subsequent segments where the atoms are the same, i.e. $\langle (u_1, b_1), \dots, (u_i, b_i), (u_{i+1}, b_{i+1}), \dots, (u_n, b_n) \rangle$ with $u_i = u_{i+1}$, then $\va(P)$ is not contractive.
On the other hand, any $P$ where $u_i \ne u_{i+1}$ is clearly repetition-proof, and $\va(P)$ will be contractive.

\begin{propositie}\label{vawssiscr}  
Let $P = \langle (u_1, b_1), \dots, (u_n, b_n) \rangle$ be a valuation path. If $u_i \ne u_{i+1}$ for all $i < n$, then $\va(P)$ is a contractive valuation algebra.
\end{propositie} 

\begin{bewijs}
Let $i \le n+1$ and $a \in \A$, then we need to show that $a/(a \bu i) = a/i$ and $a \bu a \bu i = a \bu i$ in $va(P)$. If $a \bu i = i$, then we are done. Thus assume that $a \bu i = i + 1$, in which case $i \le n$ and $u_i = a$. Since $u_{i+1} \ne u_i = a$ we get $a \bu (i + 1) = i+1$. Also, it is not hard to see that $i \le \last(a, i+1) < i+1$, but then $i = \last(a, i+1)$, and therefore $a / (i+1) = b_i = a/i$. \qed
\end{bewijs}

Based on this, our first move will be to reduce or `contract' a valuation path where some subsequent atoms are equal, to a corresponding valuation path where all subsequent atoms are different. This is not that difficult; whenever we find two subsequent segments with identical atoms, we omit one of them. More formally, we can define the contraction of a valuation path as follows:

\begin{definitie} 
Let $P$ be a valuation path. We define the \emph{contraction} of $P$, denoted $\cont(P)$, by
\begin{align*}
\cont(\e) &= \e & \cont((u, b) \cdot Q) &= (u, b) \cdot \cont_u(Q) \\
\cont_a(\e) &= \e & \cont_a((u, b) \cdot Q) &= \left\{ \begin{array}{ll}
	(u, b) \cdot \cont_u(Q) & \text{if~} u \ne a \\
	\cont_a(Q) & \text{otherwise} \\
	\end{array} \right.
\end{align*}
where $\cont_a$ is defined as above for each $a \in \A$.
\end{definitie} 

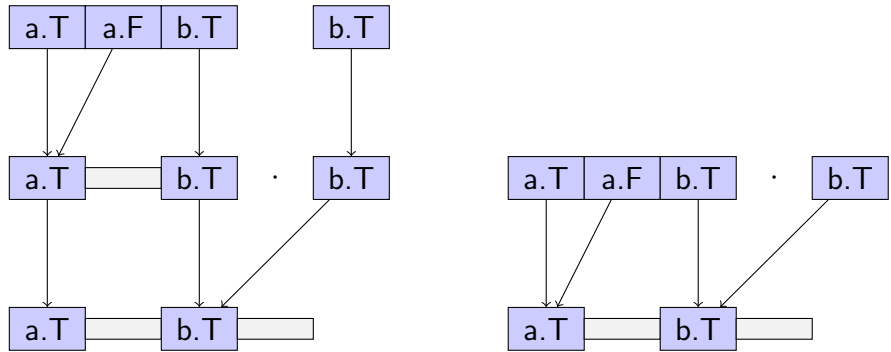
\begin{figure}
\centering
\tikzsetnextfilename{contofconcata}
\begin{tikzpicture}
\tikzstyle{pathpiece}=[draw,fill=blue!20,minimum width=10mm,font=\sffamily]
\tikzstyle{nullpiece}=[draw,fill=gray!10,minimum width=10mm,font=\sffamily]
	\node[pathpiece] (p1) {a.T};
	\node[pathpiece, right of=p1] (p2) {a.F};
	\node[pathpiece, right of=p2] (p3) {b.T};

	\node[draw=none, right of=p3] (huh) {};
	\node[pathpiece, right of=huh] (q1) {b.T};

	\node[draw=none, below of=p1] (huh) {};
	\node[pathpiece, below of=huh] (pp1) {a.T};
	\node[nullpiece, right of=pp1] (huh) {};
	\node[pathpiece, right of=huh] (pp2) {b.T};
	\path[->] (p1) edge (pp1);
	\path[->] (p2) edge (pp1);
	\path[->] (p3) edge (pp2);

	\node[draw=none, right of=pp2] (huh) {$\concat$};
	\node[pathpiece, right of=huh] (qq1) {b.T};
	\path[->] (q1) edge (qq1);

	\node[draw=none, below of=pp1] (huh) {};
	\node[pathpiece, below of=huh] (r1) {a.T};
	\node[nullpiece, right of=r1] (huh) {};
	\node[pathpiece, right of=huh] (r2) {b.T};
	\node[nullpiece, right of=r2] (huh) {};
	\path[->] (pp1) edge (r1);
	\path[->] (pp2) edge (r2);
	\path[->] (qq1) edge (r2);
\end{tikzpicture}
\hspace{10mm}
\tikzsetnextfilename{contofconcatb}
\begin{tikzpicture}
\tikzstyle{pathpiece}=[draw,fill=blue!20,minimum width=10mm,font=\sffamily]
\tikzstyle{nullpiece}=[draw,fill=gray!10,minimum width=10mm,font=\sffamily]
	\node[pathpiece] (p1) {a.T};
	\node[pathpiece, right of=p1] (p2) {a.F};
	\node[pathpiece, right of=p2] (p3) {b.T};
	\node[draw=none, right of=p3] (huh) {$\concat$};
	\node[pathpiece, right of=huh] (q1) {b.T};

	\node[draw=none, below of=p1] (huh) {};
	\node[pathpiece, below of=huh] (r1) {a.T};
	\node[nullpiece, right of=r1] (huh) {};
	\node[pathpiece, right of=huh] (r2) {b.T};
	\node[nullpiece, right of=r2] (huh) {};
	\path[->] (p1) edge (r1);
	\path[->] (p2) edge (r1);
	\path[->] (p3) edge (r2);
	\path[->] (q1) edge (r2);
\end{tikzpicture}
\caption{A schematic depiction of Example~\ref{contofconcat}.}
\label{contofconcatschem}
\end{figure}

Clearly, $\cont(\cont(P)) = \cont(P)$ for all valuation paths $P$. Example~\ref{contofconcat} shows that in general, $\cont(P \concat Q) \ne \cont(P) \concat \cont(Q)$. However, $\cont(P \concat Q) = \cont(\cont(P) \concat \cont(Q))$ for all $P$ and $Q$. These observations are illustrated by Figure~\ref{contofconcatschem}.

\begin{voorbeeld} \label{contofconcat}
For instance, let $P = \langle (a, \true), (a, \false), (b, \true) \rangle$ and $Q = \langle (b, \true) \rangle$, then $\cont(P) = \langle(a, \true), (b, \true) \rangle$, $\cont(Q) = Q$ and $\cont(P \concat Q) = \cont(\cont(P) \concat \cont(Q)) = \cont(P)$.
\end{voorbeeld}

It can be easily checked that the contraction norm defined by $||P|| = |\cont(P)|$ is a norm. We define the norm-based constructor $\vacr$ as a special case of $\va$.

\begin{definitie}\label{defvacr} 
Let $P$ be a valuation path. We define $\vacr(P) = \va(\cont(P))$.
\end{definitie} 

\begin{figure}
\centering
\tikzsetnextfilename{pathvacrpica}
\begin{tikzpicture}
\tikzstyle{pathpiece}=[draw,fill=blue!20,minimum width=10mm,font=\sffamily]
\tikzstyle{nullpiece}=[draw,fill=gray!10,minimum width=10mm,font=\sffamily]
	\node[pathpiece] 				(p1) {a.T};
	\node[pathpiece, right of=p1] (p2) {b.F};
	\node[pathpiece, right of=p2] (p3) {b.F};
	\node[pathpiece, right of=p3] (p4) {b.T};
	\node[pathpiece, right of=p4] (p5) {a.F};
	\node[pathpiece, right of=p5] (p6) {a.F};

	\node[draw=none, below of=p1] (huh) {};

	\node[pathpiece, below of=huh] (pp1) {a.T};
	\node[pathpiece, right of=pp1] (pp2) {b.F};
	\node[nullpiece, right of=pp2] (pp3) {};
	\node[nullpiece, right of=pp3] (pp4) {};
	\node[pathpiece, right of=pp4] (pp5) {a.F};
	\node[nullpiece, right of=pp5] (pp6) {};

	\path[->] (p1) edge (pp1);
	\path[->] (p2) edge (pp2);
	\path[->] (p3) edge (pp2);
	\path[->] (p4) edge (pp2);
	\path[->] (p5) edge (pp5);
	\path[->] (p6) edge (pp5);
\end{tikzpicture}
\\
\vspace{10mm}
\tikzsetnextfilename{pathvacrpicb}
\begin{tikzpicture}[->,shorten >=1pt,auto,node distance=17mm, thick]
\tikzstyle{main node}=[circle,fill=orange!20,draw,font=\sffamily\bfseries]
\tikzstyle{end node}=[circle,fill=gray!10,draw,font=\sffamily\bfseries]

  \node[main node] (1) {1};
  \node[main node] (2) [right of=1] {2};
  \node[main node] (3) [right of=2] {3};
  \node[main node] (4) [right of=3] {4};
  \node[main node] (5) [right of=4] {5};
  \node[main node] (6) [right of=5] {6};
  \node[end node] (7) [right of=6] {7};

  \path[every node/.style={font=\sffamily\small}]
    (1) edge node[above] {a.T} (2)
    (2) edge node[above] {b.F} (3)
    (3) edge node[above] {b.F} (4)
    (4) edge node[above] {b.T} (5)
    (5) edge node[above] {a.F} (6)
    (6) edge node[above] {a.F} (7)
    (7) edge[loop right] node[right] {b.T} (7)
    (1) edge[loop below] node[below] {b.F} (1)
    (2) edge[loop below] node[below] {a.T} (2)
    (3) edge[loop below] node[below] {a.T} (3)
    (4) edge[loop below] node[below] {a.T} (4)
    (5) edge[loop below] node[below] {b.T} (5)
    (6) edge[loop below] node[below] {b.T} (6)
    (7) edge[loop below] node[below] {a.F} (7);
	
  \node[main node] (11) [below =20mm of 1] {1};
  \node[main node] (22) [right of=11] {2};
  \node[draw=none] (33) [right of=22] {};
  \node[draw=none] (44) [right of=33] {};
  \node[main node] (55) [right of=44] {3};
  \node[draw=none] (66) [right of=55] {};
  \node[end node] (77) [right of=66] {4};

  \path[every node/.style={font=\sffamily\small}]
    (11) edge node[above] {a.T} (22)
    (22) edge node[above] {b.F} (55)
    (55) edge node[above] {a.F} (77)
    (77) edge[loop right] node[right] {b.F} (77)
    (11) edge[loop below] node[below] {b.F} (11)
    (22) edge[loop below] node[below] {a.T} (22)
    (55) edge[loop below] node[below] {b.F} (55)
    (77) edge[loop below] node[below] {a.F} (77);
\end{tikzpicture}
\caption{The valuation algebras $\va(P)$ and $\vacr(P)$ strongly resemble $P$ and $\cont(P)$ respectively; here, $P = \langle (a, \true), (b, \false), (b, \false), (b, \true), (a, \false), (a, \false) \rangle$.}
\label{pathvacrpic}
\end{figure}

The relation between $\va$ and $\vacr$ is illustrated by Figure~\ref{pathvacrpic}. Of course, if $\cont(P) = P$, then $\vacr(P) = \va(P)$. Proposition~\ref{vawssiscr} tells us that $\vacr$ constructs valuation algebras that are all in $\Kcn$.

Lastly, we need to construct valuation algebras that are in $\Kst$. We could again change $\va$ to do this, but here we take a simpler approach. We construct trivial valuation algebras using a norm-based constructor for the trivial norm.
By Proposition~\ref{trivialisst}, the valuation algebras constructed this way are static.

\begin{definitie}\label{defvast} 
Let $P = \langle (u_1, b_1), \dots, (u_n, b_n) \rangle$ be a valuation path. We define $\vast(P)$ as the valuation algebra $(\{1\}, /, \bu)$ where $a/1 = \true$ for $a \in \A$ if and only if there exists $i \le n$ such that $u_i = a$ and $b_i = \true$, and $a \bu 1 = 1$ for all $a \in \A$.
\end{definitie} 

\section{Satisfiability and Path-Satisfiability}
\label{sect_pathissat}

In the previous section, we have created three norm-based constructors that create valuation algebras based on valuation paths. In this section, we need to prove that these valuation algebras do what they are intended to do. That is, we need to prove that if $x$ is a formula and $P$ a valuation path such that $P \tr \se(x)$ is defined, then $x$ must evaluate to the truth value $P \tr \se(x)$ in the valuation algebra constructed by $\va$ and, under certain circumstances, also the valuation algebras constructed $\vacr$ and $\vast$.

The phrase ``under certain circumstances'' is definitely necessary. Suppose for instance that for every valuation path $P$ such that $P \tr \se(x) = \true$, $x$ evaluates to $\true$ in the valuation algebra $\vast(P)$. This would mean that every $x$ that is path-satisfiable, is satisfiable with respect to $\Kst$. If we look at Figure~\ref{prachtig2}, this results in all five satisfiabilities being the same; this is clearly not the case, as the formula $a \sand \neg a$ is satisfiable with respect to $\Kfr$, but not to $\Kst$.

The exact `circumstances' are these: $x$ must evaluate to $P \tr \se(x)$ in $\vacr(P)$ if $P$ is repetition-proof, and in $\vast(P)$ if $P$ is memorizing. The following Lemma tells us exactly what we need to prove.

\begin{lemma}\label{pathissatlemma} 
Let $x$ be a formula and $P$ a valuation path such that $P \tr \se(x)$ is defined. Then:
\begin{enumerate}
\item $x/1 = P \tr \se(x)$ in $\va(P)$;
\item if $P$ is repetition-proof, then $x/1 = P \tr \se(x)$ in $\vacr(P)$;
\item if $P$ is memorizing, then $x/1 = P \tr \se(x)$ in $\vast(P)$.
\end{enumerate}
\end{lemma} 

The proof of this lemma is based on an earlier remark: norm-based constructors such as $\va$ are somehow `invariant under concatenation'. This meant that if we take a small chunk of a valuation path, say $P_2$ as part of $P_1 \concat P_2 \concat P_3$, then the valuation algebra $\va(P_2)$ is in a sense `embedded' in $\va(P_1 \concat P_2 \concat P_3)$. In particular, this is supposed to mean that if $P_2 \tr \se(x)$ is defined, then not only does $x$ evaluate to $P_2 \tr \se(x)$ in $\va(P_2)$, but also somewhere in the larger valuation algebra $\va(P_1 \concat P_2 \concat P_3)$.

We will formalise these notions by defining that a formula is ``{\compatname}'' if it has such behaviour for all valuation paths $P_1 \concat P_2 \concat P_3$. We then proceed to prove by induction that all formulas are regular.

\begin{definitie} 
Let $\constru$ be a norm-based constructor for a norm $|| \cdot ||$. Let $\mathcal{C}$ be a collection of valuation paths. A formula $x$ is \emph{\compatname} on $\constru$ with respect to $\mathcal{C}$ if for all paths $P = P_1 \concat P_2 \concat P_3$ in $\mathcal{C}$ such that $P_2 \tr \se(x)$ is defined, the following holds:
\[
x / (||P_1|| + 1) = P_2 \tr \se(x) \text{~~and~~} x \bu (||P_1|| + 1) = ||P_1 \concat P_2|| + 1
\]
in the valuation algebra $\constru(P)$.
\end{definitie}  

As a special case, we can take $P_1 = \e = P_3$ to obtain that $x/1 = P \tr \se(x)$ in $\constru(P)$ for all $P$ in $\mathcal{C}$, for all $x$ that are {\compatname} on $\constru$ with respect to $\mathcal{C}$.
Thus, we are now left to prove that all formulas are {\compatname} on $\va$ with respect to the collection of all valuation paths, {\compatname} on $\vacr$ w.r.t. the collection of contractive paths, and {\compatname} on $\vast$ w.r.t. the collection of memorizing paths. We prove this by induction. To avoid unnecessary repetition, we use the following proposition.

\begin{propositie} \label{compatprop} 
Let $\constru$ be a norm-based constructor and let $\mathcal{C}$ be a collection of valuation paths. If every formula of the form $a$ where $a \in \A$ is {\compatname} on $\constru$ with respect to $\mathcal{C}$, then so are all other formulas.
\end{propositie} 

\begin{bewijs}
We will prove that all formulas are {\compatname} on $\constru$ with respect to $\mathcal{C}$ by induction on the complexity of the formula. Since the atoms are part of the premise, we need to consider the formulas of the form $\true$, $\neg x$ and $x \sand y$.

First, note that $\se(\true) = \seT$, thus if $P = P_1 \concat P_2 \concat P_3$ is a path in $\mathcal{C}$ with $P_2 \tr \se(\true)$ is defined, then $P_2 = \e$ and $P_2 \tr \se(\true) = \true$. As with any other valuation algebra, $\true/(||P_1|| + 1) = \true$ and $\true \bu (||P_1|| + 1) = ||P_1|| + 1 = ||P_1 \concat \e|| + 1$ in $\constru(P)$. Therefore, $\true$ is \compatname.

Let $x$ be \compatname. Let $P = P_1 \concat P_2 \concat P_3$ in $\mathcal{C}$ with $P_2 \tr \se(\neg x)$ is defined. By Proposition~\ref{resultlogic} we get that $P_2 \tr \se(\neg \neg x) = \neg (P_2 \tr \se(\neg x))$, and because $\se(\neg \neg x) = \se(x)$ we have $P_2 \tr \se(x) = \neg (P_2 \tr \se(\neg x))$. Since $x$ is \compatname, we find 
\begin{align*}
(\neg x) / (||P_1|| + 1) = \neg (x / (||P_1|| + 1)) &= \neg (P_2 \tr \se(x)) = P_2 \tr \se(\neg x), \\
(\neg x) \bu (||P_1|| + 1) = x \bu (||P_1|| + 1) &= ||P_1 \concat P_2|| + 1,
\end{align*}
in $\constru(P)$.
Thus $\neg x$ is \compatname.

Finally, let $x$ and $y$ be \compatname. Let $P = P_1 \concat P_2 \concat P_3$ in $\mathcal{C}$ with $P_2 \tr \se(x \sand y)$ is defined. If we take $X = \se(x)$, $Y = \se(y)$ and $Y' = \seF$, then we get $\se(x \sand y) = X[\seT \mapsto Y, \seF \mapsto Y']$. Now we can apply Proposition~\ref{resultdivis} to obtain paths $R$ and $Q$ such that $P_2 = R \concat Q$, $R \tr \se(x)$ is defined and
\begin{align*}
P_2 \tr \se(x \sand y) = \left\{ \begin{array}{ll}
	Q \tr \se(y) & \text{if~} R \tr \se(x) = \true \\
	Q \tr \seF & \text{otherwise}
	\end{array} \right.
\end{align*}
Note that we can write $P$ as $P_1 \concat R \concat (Q \concat P_3)$. Since $R \tr \se(x)$ is defined and $x$ is \compatname, we get that $x / (||P_1|| + 1) = R \tr \se(x)$ and $x \bu (||P_1|| + 1) = ||P_1 \concat R|| + 1$. This gives us
\begin{align*}
(x \sand y) / (||P_1|| + 1)
&= \left\{ \begin{array}{ll}
	y / (x \bu (||P_1|| + 1)) & \text{if}~ x / (||P_1|| + 1) = \true \\
	\false & \text{otherwise}
	\end{array} \right. \\
&= \left\{ \begin{array}{ll}
	y / (||P_1 \concat R|| + 1)~~ & \text{if}~ R \tr \se(x)= \true \\
	\false & \text{otherwise}
	\end{array} \right.
\end{align*}
and
\begin{align*}
(x \sand y) \bu (||P_1|| + 1)
&= \left\{ \begin{array}{ll}
	y \bu (x \bu (||P_1|| + 1)) & \text{if}~ x / (||P_1|| + 1) = \true \\
	x \bu (||P_1|| + 1) & \text{otherwise}
	\end{array} \right. \\
&= \left\{ \begin{array}{ll}
	y \bu (||P_1 \concat R|| + 1)~~ & \text{if}~ R \tr \se(x)= \true \\
	||P_1 \concat R|| + 1 & \text{otherwise}
	\end{array} \right.
\end{align*}
We will show that $(x \sand y) / (||P_1|| + 1) = P_2 \tr \se(x \sand y)$ and $(x \sand y) \bu (||P_1|| + 1) = ||P_1 \concat P_2|| + 1$ by considering both cases separately.

Suppose that $R \tr \se(x) = \true$. We can rewrite $P$ once more, this time as $(P_1 \concat R) \concat Q \concat P_3$. Because $y$ is \compatname as well and $Q \tr \se(y) = P_2 \tr \se(x \sand y)$ is defined, we now get $y / (||P_1 \concat R|| + 1) = Q \tr \se(y)$ and $y \bu (||P_1 \concat R|| + 1) = ||P_1 \concat R \concat Q|| + 1 = ||P_1 \concat P_2|| + 1$.
Suppose otherwise, then we must have $Q = \e$ since $Q \tr \false = P_2 \tr \se(x \sand y)$ is defined. Thus we get $Q \tr \false = \false$, and $||P_1 \concat R|| + 1 = ||P_1 \concat R \concat Q|| + 1$.
In either case, we have shown that $x \sand y$ is \compatname. 

This concludes the inductive proof. \qed
\end{bewijs}

We have now done a lot of hard work already.
We still need to prove that all formulas $a \in \A$ are {\compatname} with respect to the appropriate constructors and collections. We will prove each of the three parts separately, starting with $\va$.

\vspace{2mm}
\begin{bewijs}[Lemma~\ref{pathissatlemma}a]
Let $a \in \A$, then $\se(a) = \seT \sel a \ser \seF$. If $P = P_1 \concat P_2 \concat P_3$ is a path with $P_2 \tr \se(a)$ is defined, then $P_2 = \langle (a, b) \rangle$ for some $b \in \B$, and $P_2 \tr \se(a) = b$. If we write $P = \langle p_1, \dots, p_n \rangle$ and $k = |P_1| + 1$, then $p_k = (a, b)$ and $\last(a, k) = k$. From this, we get $a/(|P_1| + 1) = b_k = b$ and $a \bu (|P_1| + 1) = k+1 = |P_1| + |P_2| + 1 = |P_1 + P_2| + 1$.

Now we have shown that every $a \in \A$ is {\compatname} on $\va$ with respect to the collection of all valuation paths. By Proposition~\ref{compatprop}, every formula is {\compatname}. So, if we take a formula $x$ and a valuation path $P$ such that $P \tr \se(x)$ is defined, then we choose $P_1 = \e$, $P_2 = P$ and $P_3 = \e$ to obtain, with $||\e|| + 1 = 1$, that $x/1 = P \tr \se(x)$. \qed
\end{bewijs}
\vspace{2mm}

As $\vacr$ is based on $\va$, the proof of the second part of Lemma~\ref{pathissatlemma} is based on the previous proof. However, it is a bit more complex. 
The difficulty lies where we want to evaluate an atom $a \in \A$ that occurs in a path $P_1 \concat \langle (a, b) \rangle \concat P_3$ for some valuation paths $P_1$ and $P_3$ and some $b \in \B$. Normally, we know which valuation the atom is evaluated in, but if $P_1$ ended with $a$, this valuation will be ``contracted away'' in a sense.
For this reason, we will need a case distinction that depends on the last segment of $P_1$, and we will need to use the fact that $\vacr$ is based on contraction.

\vspace{2mm}
\begin{bewijs}[Lemma~\ref{pathissatlemma}b]
Let $a \in \A$. If $P = P_1 \concat P_2 \concat P_3$ is a repetition-proof path with $P_2 \tr \se(a)$ is defined, then $P_2 = \langle (a, b) \rangle$ for some $b \in \B$, and $P_2 \tr \se(a) = b$. We write $\cont(P) = \langle p'_1, \dots, p'_{n'} \rangle$,
$m = |\cont(P_1)|$ and $k = |\cont(P_1 \concat P_2)|$, and we note that $m \le k \le m+1$ and $k \ge 1$. Also, $\cont(P_1) = \langle p'_1, \dots, p'_m \rangle$, $\cont(P_1 \concat P_2) = \langle p'_1, \dots, p'_k \rangle$ and $p'_k = (a, b')$ for some $b' \in \B$. Note that here, $\last$ is given by the definition of $\vacr(P) = \va(\cont(P))$.

Suppose $m = 0$, then $\cont(P_1) = \e$ and $P_1 = \e$, which means $\cont(P_1 \concat P_2) = P_2$ and $k = 1$. Now we easily see $\last(a, 1) = 1$ so $a/1 = b$, and $a \bu 1 = 1+1 = |\cont(P_1 \concat P_2)| + 1$.

Suppose $m > 0$ and $u'_m \ne a$, then $\cont(P_1 \concat P_2) = \cont(P_1) \concat P_2$, which means $k = m+1$ and $p'_k = (a, b)$ so $b' = b$. We find $\last(a, m+1) = k$ thus $a/k = b'_k = b$, and $u'_k = a$ thus $a \bu k = k + 1 = |\cont(P_1 \concat P_2)| + 1$.

Suppose $m > 0$ and $u'_m = a$, then $\cont(P_1 \concat P_2) = \cont(P_1)$ and $k = m$ and $p'_m = (a, b')$. By construction $u'_{m + 1} \ne u'_m = a$, so $\last(a, m + 1) = m$, thus $a / (m + 1) = b'_m$ and $a \bu (m + 1) = m + 1 = |\cont(P_1 \concat P_2)| + 1$. By definition of $\cont(P)$, we know that $p'_m = p_i$ for some $i \le n$; that is, if we write $P_1 = \langle p_1, \dots, p_{n_1} \rangle$, $P_2 = \langle p_{n_1+1} \rangle$ and $P_3 = \langle p_{n_1 + 2}, \dots, p_n \rangle$, then there is some $i \le n_1$ such that $p_i = p'_m = (a, b')$ and $p_j = (a, b_j)$ for $i \le j \le n_1 + 1$. Because $P$ is repetition-proof, $b' = b_j$ for all $1 \le j \le n_1 + 1$, so $b' = b_{n_1+1} = b$. Now we find $a / (m+1) = b$.

Now we have shown that every $a \in \A$ is {\compatname} on $\vacr$ with respect to the collection of repetition-proof valuation paths. By Proposition~\ref{compatprop}, we are done.\qed
\end{bewijs}
\vspace{2mm}

Finally, the third part of Lemma~\ref{pathissatlemma}. This part is arguably the easiest, as the valuation algebras constructed by $\vast$ are trivial.

\vspace{2mm}
\begin{bewijs}[Lemma~\ref{pathissatlemma}c]
Let $a \in \A$. If $P = P_1 \concat P_2 \concat P_3$ is memorizing with $P_2 \tr \se(a)$ is defined, then $P_2 = \langle (a, b) \rangle$ where $b = P_2 \tr \se(a)$. If we write $P = \langle p_1, \dots, p_n \rangle$ and $k = |P_1| + 1$, then $P_2 = \langle p_k \rangle$. If $b = \true$, then $a / 1 = \true$. If not then $p_k = (a, \false)$ and there can be no other $i$ with $p_i = (a, \true)$, since $P$ is memorizing. This means $a / 1 = \false$. Also, $a \bu 1 = 1$.

Now we have shown that every $a \in \A$ is {\compatname} on $\vast$ with respect to the collection of memorizing valuation paths. By Proposition~\ref{compatprop}, we are done.\qed
\end{bewijs}
\vspace{2mm}

With Lemma~\ref{pathissatlemma} proven, we can state the following result:

\begin{stelling}\label{pathissat} 
Let $x$ be a formula.
\begin{enumerate}
\item If $\PATHSATfr(x)$, then $\SATfr(x)$.
\item If $\PATHSATrp(x)$, then $\SATcn(x)$.
\item If $\PATHSATmem(x)$, then $\SATst(x)$.
\end{enumerate}
\end{stelling} 

\begin{bewijs}
Let $P$ be a valuation path such that $P \tr \se(x) = \true$. By Lemma~\ref{pathissatlemma}a, $x/1 = \true$ in $\va(P)$. As any valuation algebra is in $\fr$, this means $\SATfr(x)$.

If $P$ is also repetition-proof (resp. memorizing), then Lemma~\ref{pathissatlemma}b (resp. \ref{pathissatlemma}c) tells us that $x/1 = \true$ in $\vacr(P)$ (resp. $\vast(P)$). By Proposition~\ref{vawssiscr} (resp. Proposition~\ref{trivialisst}), this valuation algebra is in $\cn$ (resp. $\st$), and this means $\SATcn(x)$ (resp. $\SATst(x)$). \qed
\end{bewijs}

This theorem allows us to complete our overview on the connections between satisfiability and path-satisfiability, as illustrated by Figure~\ref{prachtig3}. From this figure, another important result can be deduced: our five short-circuit logics only define three different types of satisfiability. The causes and implications of this are discussed in Chapter~\ref{ch_conclu},
but it is already shown by the following corollary.

\begin{gevolg} 
For all $x$, $\SATrp(x) \Longleftrightarrow \SATcn(x)$ and $\SATmem(x) \Longleftrightarrow \SATst(x)$.
\end{gevolg} 

\begin{bewijs}
By Proposition~\ref{geordend}, we already had one part (i.e. $\Longleftarrow$) of both equivalences.
Combining Theorem~\ref{pathissat} with Theorem~\ref{satispath} now gives us $\SATrp(x) \Longrightarrow \SATcn(x)$ and $\SATmem(x) \Longrightarrow \SATst(x)$.  \qed
\end{bewijs}

\begin{figure}
\centering
\tikzsetnextfilename{prachtig3}
\begin{tikzpicture}[->,shorten >=1pt,auto,node distance=25mm,
  main node/.style={thick, circle,fill=green!20,draw, minimum size=10pt}]

  \node[main node] (1) [label=left:{$\SATfr$}] {};
  \node[main node] (2) [label=left:{$\SATrp$}, above of=1] {};
  \node[main node] (3) [label=left:{$\SATcn$}, above of=2] {};
  \node[main node] (4) [label=left:{$\SATmem$}, above of=3] {};
  \node[main node] (5) [label=left:{$\SATst$}, above of=4] {};

  \path[dashed, every node/.style={font=\sffamily\small}]
    (5) edge (4)
    (4) edge (3)
    (3) edge (2)
    (2) edge (1)
	;

  \node[main node, fill=red!20] (11) [label=right:{$\PATHSATfr$}, right=70mm of 1] {};
  \node[main node, fill=red!20] (22) [label=right:{$\PATHSATrp$}, right=70mm of 2] {};
  \node[main node, fill=red!20] (44) [label=right:{$\PATHSATmem$}, right=70mm of 4] {};

  \path[dashed, every node/.style={font=\sffamily\small}]
    (44) edge (22)
    (22) edge (11)
	;

  \path[very thick, every node/.style={font=\sffamily\small}]
    (1) edge node[below] {$\print$} (11)
    (2) edge node[below] {$\print$} (22)
    (4) edge node[below] {$\print$} (44)
	;

  \path[very thick, every node/.style={font=\sffamily\small}]
    (11) edge[bend right=20] node[above] {$\va$} (1)
	;

  \path[very thick, every node/.style={font=\sffamily\small}]
    (22) edge node[above] {$\vacr$} (3)
    (44) edge node[above] {$\vast$} (5)
	;
\end{tikzpicture}
\caption[A third overview of satisfiability and path-satisfiability.]{An updated overview, based on Figure~\ref{prachtig2}.
The three thick arrows labeled $\va$, $\vacr$ and $\vast$ are given by Theorem~\ref{pathissat}.
}
\label{prachtig3}
\end{figure}

\chapter{Implementation in Haskell}
\label{ch_implem}

The implementation is done in Haskell, although the methods described can be adapted to most programming languages. The choice for Haskell is based mostly on ease of development, as its syntax is very reminiscent of mathematical language and therefore very suitable for satisfiability testing and theorem-proving.
Inspiration on implementing formulas in Haskell was taken from \cite{totp}.

\section{Formulas, Trees and Paths}
\label{sect_datatypes}

\subsection{Formula}

The data type Formula implements formulas.
\begin{verbatim}
data Formula    = Lit Atom
                | Const Bool
                | Neg Formula
                | Con Formula Formula
                | Dis Formula Formula
    deriving (Eq)
\end{verbatim}
The type Atom is synonymous with String. Note that this implementation considers $\false$ and $x \sor y$ to be formulas, and not abbreviations as in Chapter~\ref{ch_prelim}.
This is because the formula $\neg( \neg x \sand \neg y)$ requires more memory to store and takes more time to process. These optimisations are valued higher than the lack of redundant code.
When printed, the textual symbols \texttt{T}, \texttt{F}, \texttt{!}, \texttt{\&\&} and \texttt{||} are used to represent $\true$, $\false$, $\neg$, $\sand$ and $\sor$ respectively.

\subsection{Tree}
\label{subsect_tree}

The data type Tree implements trees.
When printed, the textual symbols \texttt{T}, \texttt{F}, \texttt{<} and \texttt{>} are used to represent $\seT$, $\seF$, $\sel$ and $\ser$ respectively.
\begin{verbatim}
data Tree   = Leaf Bool
            | Branch Tree Atom Tree
    deriving (Eq)
\end{verbatim}

Trees grow exponentially as the number of atoms grow. For each `junction', i.e. either $\sand$ or $\sor$, approximately half (either all $\true$ or all $\false$) of the leaves in the first tree are replaced by new trees. As can be shown by a simple inductive proof, the number of junctions in a constant-free formula is one less than the number of atoms in that formula. Thus, the memory size of a tree is doubled for each atom added. This means the memory required to store the $\se$-tree of a constant-free formula with $n$ atoms is estimated to be $\bigoh(2^n)$. Formulas with constants can have smaller $\se$-trees, e.g. $\se(\false \sand (a \sor b)) = \seF$. The number of leaves in the $\se$-tree of a formula with $n$ atoms is $\bigoh(2^n)$ as well.

The use of data pointers to reduce the memory requirements was considered, such as storing each branch only once and replacing leaves with pointers to branches instead of copies of trees. However, this might be more suitable for an implementation in a low level language such as C, as opposed to Haskell. Alternatively, it is possible to enumerate all possible formulas over a countably infinite set of atoms $\A$, and to use numbers to represent formulas instead of data structures. Again, however, this would be more suitable for an implementation focused on execution speed instead of experimentation and readability.

\subsection{Path}
\label{subsect_path}

Valuation paths are implemented by the type Path. Paths are simply printed as is.
\begin{verbatim}
type Path = [(Atom, Bool)]
\end{verbatim}

The function \texttt{isPathRepProof} checks if the given path is repetition-proof by recursively comparing each element of the path with the element directly after it; its complexity is $\bigoh(n)$, where $n$ is the length of the path.
\begin{verbatim}
isPathRepProof :: Path -> Bool
isPathRepProof [] = True
isPathRepProof (hd : rest) = check hd rest
  where
    check (a, b) p = case p of
        []      -> True
        (h : t) -> if (fst h) == a
            then if (snd h) == b
                then check h t
                else False
            else check h t
\end{verbatim}
The function \texttt{isPathMemorizing} checks if the given path is memorizing by keeping a list of all the bindings made; this has a worst case complexity of $\bigoh(n^2)$.
\begin{verbatim}
isPathMemorizing path = check [] path
  where
    check rs p = case p of
        []      -> True
        (h : t) -> case lookup (fst h) rs of
            (Just b)    -> if (snd h) == b
                then check rs t
                else False
            (Nothing)   -> check (h : rs) t
\end{verbatim}
The function \texttt{checkPath} chooses the appropriate function for the given logic.
\begin{verbatim}
checkPath :: Logic -> Path -> Bool
checkPath FSCL      = (\ _ -> True)
checkPath RPSCL     = isPathRepProof
checkPath CSCL      = isPathRepProof
checkPath MSCL      = isPathMemorizing
checkPath SSCL      = isPathMemorizing
\end{verbatim}

\subsection{Logic}
The data type Logic simply consists of five constants; one for each of the five short-circuit logics described in Section~\ref{sect_scl}.

\subsection{Result}
The data type Result is used by various functions as a generic piece of error-handling specific to this implementation. In particular, such functions can return \texttt{Yes}, \texttt{No} and \texttt{Unknown}. The latter is used when satisfiability testers for one logic are used on a formula in another logic, which may return render the testing inconclusive.

\subsection{Normal Form}
\label{subsect_normalformmake}
As mentioned in Section~\ref{sect_normalform}, a function $f$ exists which maps formulas to normal form equivalents.
In Section~\ref{sect_normalform}, it is discussed that normal forms resemble $\se$-trees. In Section~\ref{sect_pathsat}, this is expanded upon by two corollaries, Corollary~\ref{normalformpathsat} and Corollary~\ref{normalformpathsat2}, that suggest that the function $f$ can be used to determine path-satisfiability. Unfortunately, just like $\se$-trees (Section~\ref{subsect_tree}), the normal forms grow exponentially in size; for a formula containing $n$ atoms, which is thus of size $\bigoh(n)$, the normal form formula created by applying $f$ is of size $\bigoh(2^n)$.

\section{Satisfiability Testers}
\label{sect_sattesters}

The data type SatTester implements satisfiability testers for short-circuit logic.
\begin{verbatim}
type SatTester = Logic -> Formula -> Result
\end{verbatim}
A SatTester is a function that determines if a formula $x$ is path-satisfiable with regards to a certain logic. A SatTester should result either \texttt{Yes p} if \texttt{p} is a path that satisfies the formula, \texttt{No} if the formula is not path-satisfiable, or \texttt{Unknown} if it is not able to conclude either answer with certainty.
Five SatTesters are implemented: SatBruteControl, SatBruteForce, SatDirect, SatOpen, and SatBoolean.

\subsection{SatBruteControl}

SatBruteControl is a $\bigoh(2^n)$ satisfiability tester for all logics.
It tries to construct a path based on the $\se$-tree of the formula, by searching the leaves for $\seT$. If it is found, the path created along the way is checked using the \texttt{checkPath} function.
\begin{verbatim}
findValidSolution :: Logic -> Tree -> Path -> (Bool, Path)
findValidSolution lg et p = case et of
    (Leaf True)     -> (checkPath lg p, p)
    (Leaf False)    -> (False, p)
    (Branch l c r)  ->
        let
          soll = findValidSolution lg l (p ++ [(c, True)])
          solr = findValidSolution lg r (p ++ [(c, False)])
        in if fst soll
            then (fst soll, (c, True) : (snd soll))
            else (fst solr, (c, False) : (snd solr))
\end{verbatim}
SatBruteControl continues searching until an appropriate $\seT$ leaf is found, thus in the worst case this involves building and searching the entire tree, which is a $\bigoh(2^n)$ operation. Each time a leaf is found, the constructed path has to be checked, which is a mere $\bigoh(n^2)$ operation in the worst case (see Section~\ref{subsect_path}).

\subsection{SatBruteForce}

SatBruteForce is also a $\bigoh(2^n)$ satisfiability tester for all logics.
It is a minor improvement over SatBruteControl. For MSCL and SSCL they coincide. For FSCL, the path is not checked since any path is valid, which eliminates the need to carry the path down the recursion.
\begin{verbatim}
findAnySolution :: Tree -> (Bool, Path)
findAnySolution et = case et of
    (Leaf True)     -> (True, [])
    (Leaf False)    -> (False, [])
    (Branch l c r)  ->
        let
          soll = findAnySolution l
          solr = findAnySolution r
        in if fst soll
            then (fst soll, (c, True) : (snd soll))
            else (fst solr, (c, False) : (snd solr))
\end{verbatim}
For RPSCL and CSCL, the path that is created while searching for $\seT$ leaves is directly checked to be repetition-proof by carrying the last encountered atom down the recursion.
\begin{verbatim}
findRepProofSolution :: Tree -> Maybe (Atom, Bool) -> (Bool, Path)
findRepProofSolution et m = case et of
    (Leaf True)     -> (True, [])
    (Leaf False)    -> (False, [])
    (Branch l c r)  -> case m of
        (Nothing)       ->
            let
              soll = findRepProofSolution l (Just (c, True))
              solr = findRepProofSolution r (Just (c, False))
            in if fst soll
                then (fst soll, (c, True) : (snd soll))
                else (fst solr, (c, False) : (snd solr))
        (Just (a, b))   -> if (a == c)
            then
                let
                  p = (if b then l else r)
                  solp = findRepProofSolution p (Just (c, b))
                in (fst solp, (c, b) : (snd solp))
            else findRepProofSolution et (Nothing)
\end{verbatim}
Both methods still require most of the $\se$-tree to be searched, so SatBruteForce is $\bigoh(2^n)$ in the worst case for all logics.

\subsection{SatDirect}

SatDirect is a $\bigoh(n)$ satisfiability tester for FSCL, but can be used for other logics as well. If no path is found, then this means the formula is not satisfiable for any logic. If a path is found, then either the path is usable within the logic and \texttt{Yes} is returned, or it is not, in which case \texttt{Unknown} is returned.

It is based on remarks made in \cite{propalg}, which state
\begin{align*}
\SATfr(\true)& &\neg \FALfr(\true)& \\
\SATfr(a)& &\FALfr(a)& \\
\SATfr(\neg x) &\Leftrightarrow \FALfr(x) & \FALfr(\neg x) &\Leftrightarrow \SATfr(x) \\
\SATfr(x \sand y) &\Leftrightarrow \SATfr(x) \wedge \SATfr(y) & \FALfr(x \sand y) &\Leftrightarrow \FALfr(x) \vee \FALfr(y)
\end{align*}
where $a$ ranges over $\A$. This allows us to tell if a formula is satisfiable or not.

The function \texttt{sat} determines if a formula is satisfiable, and supplies a satisfying path if it is. If it is not, the supplied path is discarded.
\begin{verbatim}
sat :: Formula -> (Bool, Path)
sat (Const True)        = (True, [])
sat (Const False)       = (False, [])
sat (Lit a)             = (True, [(a, True)])
sat (Neg f)             = fal f
sat (Con f1 f2)         = if fst (sat f1)
    then (fst (sat f2), snd (sat f1) ++ snd (sat f2))
    else sat f1
sat (Dis f1 f2)         = if fst (sat f1)
    then sat f1
    else (fst (sat f2), snd (sat f1) ++ snd (sat f2))
\end{verbatim}
The definition of \texttt{fal} is similar, but the roles of $\true$ and $\false$, and of $\sand$ and $\sor$ are swapped; it determines if a formula is falsifiable instead.

Each atom is only visited once. As FSCL imposes no restrictions, the decision at each atom is trivial: `true' if we want the formula to be satisfiable, or `false' if we want it to be falsifiable. Note that for a formula $x \sand y$, or \texttt{Con x y} in the code, we first try satisfiability for $x$. If $x$ is satisfiable, we concatenate this result with the result of $y$. If it is not, then $x \sand y$ cannot be satisfiable, so we are done. In this manner, SatDirect itself uses short-circuit evaluation to determine satisfiability.

\subsection{SatOpen}

SatOpen is a $\bigoh(n)$ satisfiability tester for RPSCL and CSCL, but can be used for other logics as well. If no path is found, then \texttt{No} is returned for logics other than FSCL. If a path is found, then either \texttt{Yes} or \texttt{Unknown} is returned based on the usability within the given logic.

It is based on SatDirect. Again, each atom only needs to be visited once, but this time the decision to make an atom either `true' or `false' is more complicated.
As an example, a naive way to try to solve $\SATrp((a \sor b) \sand \neg a)$ would do the following: first, we examine $a \sor b$. This can be made true by taking $a$ to be true. Now, we proceed to $\neg a$, and discover that $a$ must be false immediately afterwards. This is not allowed under the rules and regulations of rp-path-satisfiability, so we might wrongly assume that $(a \sor b) \sand \neg a$ is not rp-path-satisfiable. Of course we can take the path $\langle (a, \false), (b, \true), (a, \false) \rangle$ and show that it is even mem-path-satisfiable. The decision made when we first came across $a$ must have been the wrong one, but this seems impossible to tell without knowing what lies ahead. This lack of knowledge of future events is one of the main obstacles in solving propositional satisfiability, where guessing (and backtracking in case the guess was wrong) is usually the only option.

In the case of short-circuit logic, the solution is much simpler: we simply work from the back to the front. The formula $\neg a$ is satisfiable, but only by taking $a$ to be false. With this knowledge we try to make $a \sor b$ satisfiable. If we encounter $a$, then we know we must take $a$ to be false, preventing the creation of paths that are not repetition-proof. In the code, this knowledge is represented by a `guard':
\begin{verbatim}
type Guard = (Maybe Atom, Maybe Path, Maybe Path)
\end{verbatim}
A guard is either empty (if no atoms have been assigned a value) or it contains an atom. If it does, then it must contain either a path where that atom is true, one where that atom is false, or both. 
These guards are used when determining if an atom is satisfiable.
\begin{verbatim}
(Lit a) -> case g of
    (Just _, Just p, _)             
        -> (Just (Just a, Just ((a, True) : p), Nothing))
    (Just u, Nothing, Just p)       
        -> if u == a
            then (Nothing)
            else (Just (Just a, Just ((a, True) : p), Nothing))
    (Nothing, Nothing, Nothing)     
        -> (Just (Just a, Just [(a, True)], Nothing))
    _                               
        -> error ("illegal u-guard" ++ show g)
\end{verbatim}
If the guard is empty, contains a different atom, or contains a path starting with $(a, \true)$, the atom $a$ can be made true without a problem. If the guard `forces' $a$ to be false, then it cannot also be true, so the formula is not satisfiable.
Once a new atom is assigned a value, a new guard is made, and the old guard can be discarded. This allows the algorithm to remain $\bigoh(n)$.

In the introduction to this thesis, we explained that a formula $x \sand y$ can only be true if both $x$ and $y$ are true, and that knowing $x$ is false is enough to determine that $x \sand y$ is also false. But of course the same holds if we know that $y$ is false. If we want to know if $x \sand y$ is rp-path-satisfiable, we first determine if $y$ is. If it is not, then we do not need to consider the satisfiability of $x$. In this way, SatOpen uses short-circuit evaluation as well, but right-sequentially instead of left-sequentially. 

\subsection{SatBoolean}
\label{subsect_satboolean}
SatBoolean is a satisfiability tester for MSCL and SSCL.

\begin{figure}
\centering
\tikzsetnextfilename{prachtig4}
\begin{tikzpicture}[->,shorten >=1pt,auto,node distance=25mm,
  main node/.style={thick, circle,fill=green!20,draw, minimum size=10pt}]

  \node[main node] (1) [label=left:{$\SATfr$}] {};
  \node[main node] (2) [label=left:{$\SATrp$}, above of=1] {};
  \node[main node] (3) [label=left:{$\SATcn$}, above of=2] {};
  \node[main node] (4) [label=left:{$\SATmem$}, above of=3] {};
  \node[main node] (5) [label=left:{$\SATst$}, above of=4] {};

  \path[dashed, every node/.style={font=\sffamily\small}]
    (5) edge (4)
    (4) edge (3)
    (3) edge (2)
    (2) edge (1)
	;

  \node[main node, fill=red!20] (11) [label=right:{$\PATHSATfr$}, right=70mm of 1] {};
  \node[main node, fill=red!20] (22) [label=right:{$\PATHSATrp$}, right=70mm of 2] {};
  \node[main node, fill=red!20] (44) [label=right:{$\PATHSATmem$}, right=70mm of 4] {};
  \node[main node, fill=blue!20] (55) [label=right:{$\PROPSAT$}, right=70mm of 5] {};

  \path[dashed, every node/.style={font=\sffamily\small}]
    (44) edge (22)
    (22) edge (11)
	;

  \path[dashed, every node/.style={font=\sffamily\small}]
    (5) edge (55)
    (55) edge[bend right=20] (5)
	;

  \path[very thick, every node/.style={font=\sffamily\small}]
    (55) edge node[right] {\texttt{makePath}} (44)
	;

  \path[very thick, every node/.style={font=\sffamily\small}]
    (1) edge node[below] {$\print$} (11)
    (2) edge node[below] {$\print$} (22)
    (4) edge node[below] {$\print$} (44)
	;

  \path[very thick, every node/.style={font=\sffamily\small}]
    (11) edge[bend right=20] node[above] {$\va$} (1)
	;

  \path[very thick, every node/.style={font=\sffamily\small}]
    (22) edge node[above] {$\vacr$} (3)
    (44) edge node[above] {$\vast$} (5)
	;
\end{tikzpicture}
\caption[A fourth overview of satisfiability and path-satisfiability.]{A final overview, based on Figure~\ref{prachtig3}.
The blue node represents propositional satisfiability, as mentioned in Section~\ref{sect_scl} and Section~\ref{sect_satis}.
The two dashed arrows connected to it are given by Corollary~\ref{satssclisbinsat}. The thick arrow labeled \texttt{makePath}
leading down from it is discussed in Section~\ref{subsect_satboolean}.
}
\label{prachtig4}
\end{figure}
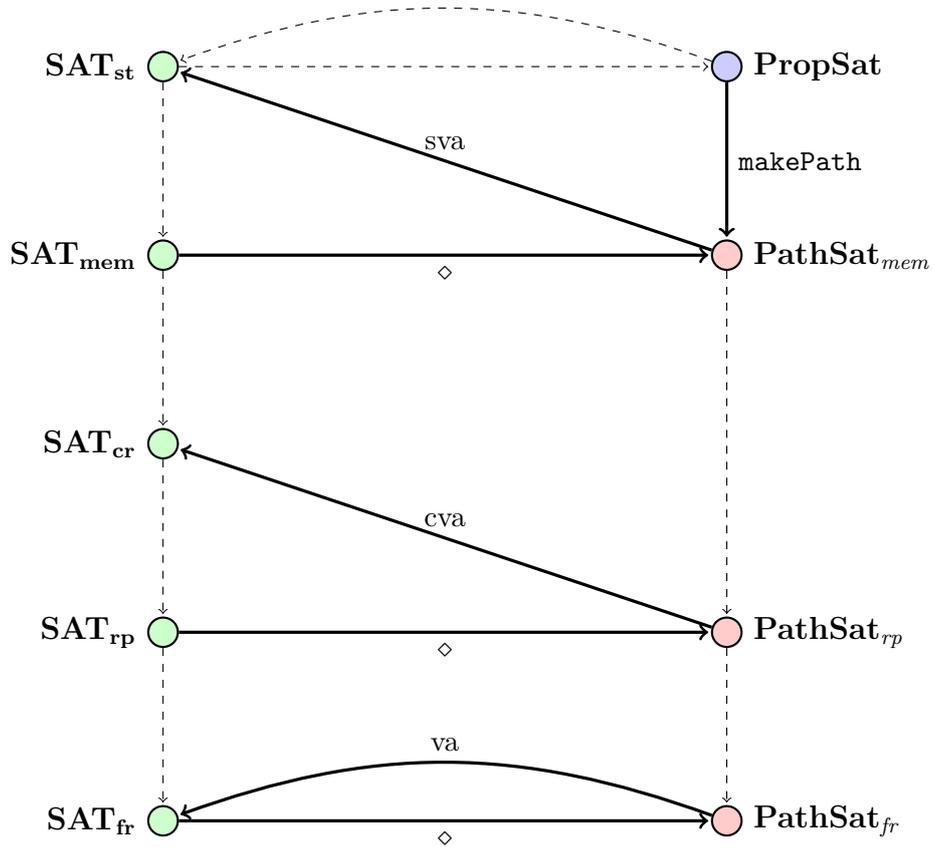

In Section~\ref{sect_satis}, it is discussed that satisfiability for SSCL coincides with propositional satisfiability. 
See Figure~\ref{prachtig4} for an overview of this.
Propositional satisfiability, also called boolean satisfiability, has already been the subject of many research papers, and has been proven to be NP-complete by Stephen Cook in 1971, as discussed in \cite{cook} and \cite{schaefer};
therefore, a new implementation would be mostly pointless.
SatBoolean instead provides a wrapper function for a boolean satisfiability solver, an implementation of propositional satisfiability that uses the Davis-Putnam-Logemann-Loveland algorithm defined in \cite{dpll1} and \cite{dpll2}. Its complexity is the same as that of the DPLL algorithm, thus $\bigoh(2^n)$ worst case.

First, the formula is translated to the correct format. Then the imported SatSolver module is used to determine propositional satisfiability. If no solution is found, \texttt{No} is returned for the logics MSCL and SSCL; for other logics, \texttt{Unknown} is returned. If a solution is found, the function \texttt{makePath} uses it to construct a valuation path.

\chapter{Conclusion}
\label{ch_conclu}

This thesis set out to define evaluation and satisfiability for each of the five short-circuit logics defined in \cite{scl1}; FSCL (`free short-circuit logic'), RPSCL (`repetition-proof'), CSCL (`contractive'), MSCL (`memorizing') and SSCL (`static'). 
The desire to implement a program that could test this satisfiability lead to a different definition based on paths, which we called `path-satisfiability'.
Three types of valuation paths were defined, corresponding to the terms `free', `repetition-proof' and `memorizing'.

A considerable portion of this thesis was spent proving the correspondences between the theoretically defined satisfiability and the path-satisfiability that was implemented. 
The result of this work was that what had appeared to be five types of satisfiability, one for each short-circuit logic, turned out to be only three types.
It is proven that RPSCL and CSCL generate the same form of satisfiability, which is more restricted than that of FSCL, but less so than that of MSCL. 
The semantics of `repetition-proof' restrict what truth values certain atoms can take, whereas the semantics of `contractive' restrict what side-effects they can have. 
In the context of satisfiability, these side-effects only serve to alter truth values, so at that point the second restriction placed by CSCL is moot.
A similar situation occurs between MSCL and SSCL; 
everything that can be achieved by using side-effects that obey the laws of MSCL, can also be achieved without the use of any side-effects.
They too are proven to share their satisfiability.

In \cite{scl1} and \cite{propalg}, it was already mentioned that static short-circuit logic was a sequential version of propositional logic. 
This was further discussed in this thesis; satisfiability for MSCL and SSCL both turned out to be equivalent to propositional (or `boolean') satisfiability. 
Furthermore, it was shown that satisfiability for FSCL could be solved by using the short-circuit behaviour of the connectives $\sand$ and $\sor$; 
satisfiability for RPSCL and CSCL could as well, but here this behaviour was right-sequential, as we worked from right to left instead.

\vspace{5mm}
Our implementation was more experimental in nature, and has room for many kinds of improvements. 
A reimplementation of the algorithms described in Chapter~\ref{ch_implem} could lead to more memory and time efficiency, thus to a more practical program. 
For analysing especially large formulas, such a new implementation could use parallelisation and even memoization to avoid doing double work.

Additionally, in the introduction it was mentioned that satisfiability for short-circuit logic could be applied to dead code detection. This might be something worth investigating in future papers.




\appendix

\chapter{Axioms of Short Circuit Logics}
\label{ch_axioms}

\begin{opmerking}
Note that because $\false$ and $\sor$ are defined as abbreviations in this thesis, axioms \ref{eqn:deffalse} and \ref{eqn:defsor} are technically not axioms but defining equations.
\end{opmerking}

The system $\EqFSCL$ consists of the following 10 axioms:
\begin{align}
	\label{eqn:deffalse}
\false &= \neg \true \\
	\label{eqn:defsor}
x \sor y &= \neg (\neg x \sand \neg y) \\
	\label{eqn:negneg}
\neg \neg x &= x \\
	\label{eqn:tandx}
\true \sand x &= x \\
	\label{eqn:xandt}
x \sand \true &= x \\
	\label{eqn:fandx}
\false \sand x &= \false \\
	\label{eqn:commut}
(x \sand y) \sand z &= x \sand (y \sand z) \\
	\label{eqn:xandf}
x \sand \false &= \neg x \sand \false \\
	\label{eqn:righths}
(x \sand \false) \sor y &= (x \sor \true) \sand y \\
	\label{eqn:distri}
(x \sand y) \sor (z \sand \false) &= (x \sor (z \sand \false)) \sand (y \sor (z \sand \false))
\end{align}

The system $\EqRPSCL$ extends $\EqFSCL$ with the following axiom schemes, where $a$ ranges over $\A$:
\begin{align}
a \sand (a \sor x) &= a \sand a \\
a \sor (a \sand x) &= a \sand a \\
(a \sor \neg a) \sand x &= (\neg a \sand a) \sor x \\
(\neg a \sor a) \sand x &= (a \sand \neg a) \sor x \\
(a \sand \neg a) \sand x &= a \sand \neg a \\
(\neg a \sand a) \sand x &= \neg a \sand a \\
(x \sand y) \sor (a \sand \neg a) &= (x \sor (a \sand \neg a)) \sand (y \sor (a \sand \neg a)) \\
(x \sand y) \sor (\neg a \sand a) &= (x \sor (\neg a \sand a)) \sand (y \sor (\neg a \sand a))
\end{align}

The system $\EqCSCL$ extends $\EqFSCL$ with the following axiom schemes, where $a$ ranges over $\A$:
\begin{align}
a \sand (a \sor x) &= a \\
a \sor (a \sand x) &= a \\
a \sor \neg a &= a \sor \true \\
a \sand \neg a &= a \sand \false
\end{align}

The system $\EqMSCL$ is based on $\EqFSCL$ but replaces axioms \ref{eqn:righths} and \ref{eqn:distri} with the following axioms:
\begin{align}
x \sand (x \sor y) &= x \\
x \sand (y \sor z) &= (x \sand y) \sor (x \sand z) \\
(x \sand y) \sor (\neg x \sand z) &= (x \sor z) \sand (\neg x \sor y) \\
(x \sand y) \sor (\neg x \sand z) &= (\neg x \sand z) \sor (x \sand y) \\
((x \sand y) \sor (\neg x \sand z)) \sand u &= (x \sand (y \sand u)) \sor (\neg x \sand (z \sand u))
\end{align}

Finally, the system $\EqSSCL$ extends $\EqMSCL$ with one final axiom:
\begin{align}
x \sand \false &= \false
\end{align}

\chapter{Proof of Proposition~\ref{memensthandig}}
\label{ch_proofs}

\begin{propositie} \label{prophalfsupermem} 
Let $V$ be a valuation algebra. If $V$ is memorizing and $a \in \A$ then
\begin{align}\label{halfsupermem}
a/(x \bu a \bu H) &= a/H, &
a \bu x \bu a \bu H &= x \bu a \bu H
&& (\forall H \in V)
\end{align}
for all formulas $x$.
\end{propositie} 

\begin{bewijs}
Let $V$ be memorizing and let $a \in \A$. Because $V$ is also contractive and repetition-proof, we find 
\begin{align*}
a/(\true \bu a \bu H) = a/(a \bu H) = a/H, &&
a \bu \true \bu a \bu H = a \bu a \bu H = a \bu H
\end{align*}
for all $H \in V$, so $\true$ satisfies (\ref{halfsupermem}). Let $b \in \A$, then $b$ satisfies (\ref{halfsupermem}) because $V$ is memorizing. Suppose $x, y$ are formulas that satisfy (\ref{halfsupermem}).
Because $(\neg x) \bu a \bu H = x \bu a \bu H$ for all $H$, it immediately follows that $\neg x$ also satisfies (\ref{halfsupermem}). Additionally, because 
\begin{align*}
a/(y \bu x \bu a \bu H)
&= a/(y \bu (x \bu a \bu H)) \\
&= a/(y \bu (a \bu x \bu a \bu H)) \\
&= a/(y \bu a \bu (x \bu a \bu H)) \\
&= a/(x \bu a \bu H) \\
&= a/H
\end{align*}
and
\begin{align*}
a \bu y \bu x \bu a \bu H
&= a \bu y \bu (x \bu a \bu H) \\
&= a \bu y \bu (a \bu x \bu a \bu H) \\
&= a \bu y \bu a \bu (x \bu a \bu H) \\
&= y \bu a \bu (x \bu a \bu H) \\
&= y \bu (a \bu x \bu a \bu H) \\
&= y \bu (x \bu a \bu H) \\
&= y \bu x \bu a \bu H,
\end{align*}
we find
\begin{align*}
a/((x \sand y) \bu a \bu H)
&= 	\left\{ \begin{array}{ll}
	a/(x \bu a \bu H) 									& \text{if~} x/(a \bu H) = \false \\
	a/(y \bu x \bu a \bu H)							& \text{otherwise}
	\end{array} \right. \\
&= 	\left\{ \begin{array}{ll}
	a/H			\phantom{(y\bu x\bu a\bu{})} 			& \text{if~} x/(a \bu H) = \false \\
	a/H 													& \text{otherwise}
	\end{array} \right. \\
&= a/H
\end{align*}
and
\begin{align*}
a \bu (x \sand y) \bu a \bu H
&= 	\left\{ \begin{array}{ll}
	a \bu x \bu a \bu H 									& \text{if~} x/(a \bu H) = \false \\
	a \bu y \bu x \bu a \bu H								& \text{otherwise}
	\end{array} \right. \\
&= 	\left\{ \begin{array}{ll}
	x \bu a \bu H											& \text{if~} x/(a \bu H) = \false \\
	y \bu x \bu a \bu H		\phantom{a \bu {}}			& \text{otherwise}
	\end{array} \right. \\
&= (x \sand y) \bu a \bu H.
\end{align*}
This means also $x \sand y$ satisfies (\ref{halfsupermem}). By induction, all formulas $x$ satisfy (\ref{halfsupermem}). \qed
\end{bewijs}

\begin{propositie} \label{prophalfsuperst} 
Let $V$ be a valuation algebra. If $V$ is static and $a \in \A$ then
\begin{align}\label{halfsuperst}
a/(x \bu H) &= a/H 
&& (\forall H \in V)
\end{align}
for all formulas $x$.
\end{propositie} 

\begin{bewijs}
Let $V$ be static and let $a \in \A$. Clearly, $a/(\true \bu H) = a/H$ for all $H \in V$, so $\true$ satisfies (\ref{halfsuperst}). Also, let $b \in \A$, then $b$ satisfies (\ref{halfsuperst}) because $V$ is static. Suppose $x, y$ are formulas that satisfy (\ref{halfsuperst}). Then $a/((\neg x) \bu H) = a/(x \bu H) = a/H$, so $\neg x$ satisfies (\ref{halfsuperst}). Furthermore, because $a/(y \bu (x \bu H)) = a/(x \bu H) = a/H$ we find
\begin{align*}
a/((x \sand y) \bu H)
&= 	\left\{ \begin{array}{ll}
	a/(x \bu H) 						& \text{if~} x/H = \false \\
	a/(y \bu x \bu H) 					& \text{otherwise}
	\end{array} \right. \\
&= 	\left\{ \begin{array}{ll}
	a/H 		\phantom{(y\bu x\bu{})} 	& \text{if~} x/H = \false \\
	a/H  							& \text{otherwise}
	\end{array} \right. \\
&= a/H.
\end{align*}
and this means $x \sand y$ satisfies (\ref{halfsuperst}). By induction, we are done. \qed
\end{bewijs}

\begin{propositie*}[\ref{memensthandig}a] 
Let $V$ be a valuation algebra. If $V$ is memorizing then
\begin{align}\label{supermem}
x/(y \bu x \bu H) &= x/H, &
x \bu y \bu x \bu H &= y \bu x \bu H
&& (\forall H \in V)
\end{align}
for all formulas $x$, $y$.
\end{propositie*} 

\begin{bewijs}
Let $V$ be memorizing and fix a formula $y$. We will prove this proposition by induction to the complexity of $x$. The $\true$ case is immediate. The $a$ case for $a \in \A$ is already given by Proposition~\ref{prophalfsupermem}.

The case $x = \neg x_1$ where $x_1$ satisfies (\ref{supermem}): we can derive
\begin{align*}
(\neg &x_1) / (y \bu (\neg x_1) \bu H) & (\neg &x_1) \bu y \bu (\neg x_1) \bu H \\
&= \neg (x_1 / (y \bu x_1 \bu H)) & &= x_1 \bu y \bu x_1 \bu H \\
&= \neg (x_1 / H) & &= y \bu x_1 \bu H \\
&= (\neg x_1) / H & &= y \bu (\neg x_1) \bu H
\end{align*}
and therefore $\neg x_1$ also satisfies (\ref{supermem}).

For the case $x = x_1 \sand x_2$ where $x_1$ and $x_2$ both satisfy (\ref{supermem}), we will consider two possibilities separately:

Suppose $x_1 / H = \true$, then $(x_1 \sand x_2) / H = x_2 / (x_1 \bu H)$ and $(x_1 \sand x_2) \bu H = x_2 \bu x_1 \bu H$. We will use a rewriting trick:
if $u$ and $v$ are formulas and $H$ is a valuation, then
\begin{align*}
u \bu v \bu H &= ((v \sor \true) \sand u) \bu H.
\end{align*}
Now we can derive
\begin{align*}
(x_1 &\sand x_2) / (y \bu (x_1 \sand x_2) \bu H) \\
&= \left\{ \begin{array}{ll}
  x_2 / (x_1 \bu y \bu x_2 \bu x_1 \bu H) \hspace{20mm}
      & \text{if~} x_1 / (y \bu x_2 \bu x_1 \bu H) = \true \\
  \false 
      & \text{otherwise}
  \end{array} \right. \\
&= \left\{ \begin{array}{ll}
  x_2 / (((y \sor \true) \sand x_1) \bu x_2 \bu (x_1 \bu H))
      & \text{if~} x_1 / (((x_2 \sor \true) \sand y) \bu x_1 \bu H) = \true \\
  \false 
      & \text{otherwise}
  \end{array} \right. \\
&= \left\{ \begin{array}{ll}
  x_2 / (x_1 \bu H) \hspace{42.5mm} 
      & \text{if~} x_1 / H = \true \\
  \false 
      & \text{otherwise}
  \end{array} \right. \\
&= (x_1 \sand x_2) / H
\end{align*}
and
\begin{align*}
(x_1 &\sand x_2) \bu y \bu (x_1 \sand x_2) \bu H \\
&= \left\{ \begin{array}{ll}
  x_2 \bu x_1 \bu y \bu x_2 \bu x_1 \bu H \hspace{20mm}
      & \text{if~} x_1 / (y \bu x_2 \bu x_1 \bu H) = \true \\
  x_1 \bu y \bu x_2 \bu x_1 \bu H 
      & \text{otherwise}
  \end{array} \right. \\
&= \left\{ \begin{array}{ll}
  x_2 \bu x_1 \bu y \bu x_2 \bu x_1 \bu H \hspace{20mm}
      & \text{if~} x_1 / H = \true \\
  x_1 \bu y \bu x_2 \bu x_1 \bu H 
      & \text{otherwise}
  \end{array} \right. \\
&= x_2 \bu x_1 \bu y \bu x_2 \bu x_1 \bu H \\
&= x_2 \bu ((y \sor \true) \sand x_1) \bu x_2 \bu (x_1 \bu H) \\
&= ((y \sor \true) \sand x_1) \bu x_2 \bu (x_1 \bu H) \\
&= x_1 \bu y \bu x_2 \bu x_1 \bu H \\
&= x_1 \bu ((x_2 \sor \true) \sand y) \bu x_1 \bu H \\
&= ((x_2 \sor \true) \sand y) \bu x_1 \bu H \\
&= y \bu x_2 \bu x_1 \bu H \\
&= y \bu (x_1 \sand x_2) \bu H
\end{align*}
so it satisfies (\ref{supermem}).

Suppose otherwise, i.e. $x_1 / H = \false$, then $(x_1 \sand x_2) / H = \false$ and $(x_1 \sand x_2) \bu H = x_1 \bu H$. We can derive
\begin{align*}
(x_1 \sand x_2) / (y \bu (x_1 \sand x_2) \bu H)
&= (x_1 \sand x_2) / (y \bu x_1 \bu H) \\
&= \left\{ \begin{array}{ll}
  x_2 / (y \bu x_1 \bu H) 
      & \text{if~} x_1 / (y \bu x_1 \bu H) = \true \\
  \false 
      & \text{otherwise}
  \end{array} \right. \\
&= \left\{ \begin{array}{ll}
  x_2 / (y \bu x_1 \bu H) 
      & \text{if~} x_1 / H = \true \\
  \false 
      & \text{otherwise}
  \end{array} \right. \\
&= \false
\end{align*}
and
\begin{align*}
(x_1 \sand x_2) \bu y \bu (x_1 \sand x_2) \bu H
&= (x_1 \sand x_2) \bu y \bu x_1 \bu H \\
&= \left\{ \begin{array}{ll}
  x_2 \bu x_1 \bu y \bu x_1 \bu H 
      & \text{if~} x_1 / (y \bu x_1 \bu H) = \true \\
  x_1 \bu y \bu x_1 \bu H 
      & \text{otherwise}
  \end{array} \right. \\
&= \left\{ \begin{array}{ll}
  x_2 \bu x_1 \bu y \bu x_1 \bu H 
      & \text{if~} x_1 / H = \true \\
  x_1 \bu y \bu x_1 \bu H 
      & \text{otherwise}
  \end{array} \right. \\
&= x_1 \bu y \bu x_1 \bu H \\
&= y \bu x_1 \bu H \\
&= y \bu (x_1 \sand x_2) \bu H
\end{align*}
so now it satisfies (\ref{supermem}) as well.

Thus the case $x_1 \sand x_2$ also satisfies (\ref{supermem}), which concludes our inductive proof. \qed
\end{bewijs}

\begin{propositie*}[\ref{memensthandig}b] 
Let $V$ be a valuation algebra.
If $V$ is static then 
\begin{align} \label{superst}
x/(y \bu H) = x/H
&& (\forall H \in V)
\end{align}
for all formulas $x$, $y$.
\end{propositie*} 

\begin{bewijs}
Let $V$ be static and fix a formula $y$. We will prove this proposition by induction to the complexity of $x$. The $\true$ case is immediate. The $a$ case for $a \in \A$ is already given by Proposition~\ref{prophalfsuperst}.

The case $x = \neg x_1$ where $x_1$ satisfies (\ref{superst}): we can derive
\begin{align*}
(\neg x_1) / (y \bu H) 
= \neg (x_1 / (y \bu H)) 
= \neg (x_1 / H) 
= (\neg x_1) / H
\end{align*}
and therefore $\neg x_1$ also satisfies (\ref{superst}).

The case $x = x_1 \sand x_2$ where $x_1$ and $x_2$ both satisfy (\ref{superst}): we derive $x_1 / (y \bu H) = x_1 / H$ and $x_2 / (x_1 \bu (y \bu H)) = x_2 / (y \bu H) = x_2 / H$, thus
\begin{align*}
(x_1 \sand x_2) / (y \bu H)
&= \left\{ \begin{array}{ll}
  x_2 / (x_1 \bu y \bu H) & \text{if~} x_1 / (y \bu H) = \true \\
  \false & \text{otherwise}
  \end{array} \right. \\
&= \left\{ \begin{array}{ll}
  x_2 / H \phantom{(x_1 \bu y \bu {})} & \text{if~} x_1 / H = \true \\
  \false & \text{otherwise}
  \end{array} \right. \\
&= (x_1 \sand x_2) / H
\end{align*}
and therefore $x_1 \sand x_2$ also satisfies (\ref{superst}). \qed
\end{bewijs}

\end{document}